\documentclass[aps,prb,reprint]{revtex4-1}

\usepackage{amsmath} 
\usepackage{graphics} 
\usepackage{graphicx} 
\usepackage{caption} 
\usepackage{array} 
\usepackage{multirow} 

\begin{document}
\def\useministyle#1${\scriptstyle{#1}$}
\def\Matrix#1{{\everymath{\useministyle}\bordermatrix{#1}}}
\def\sss{\scriptstyle}
\def\erule{\vrule height 0pt depth 0pt width 0pt}

\title{The Ising model in planar lacunary and fractal lattices, a path counting approach}
\date{\today}
\author{Michel Perreau}
\affiliation{Universit\'e Paris Diderot, Sorbonne Paris Cit\'e, 5 rue Thomas Mann, 75205 Paris Cedex 13, France}
\email{michel.perreau@univ-paris-diderot.fr}

\begin{abstract}
The method of counting loops for calculating the partition function of the Ising model on the two dimensional square lattice is extended to lacunary planar lattices, especially scale invariant fractal lattices, the Sierpi\'nsky carpets with different values of the scale invariance ratio and of the number of deleted sites. The critical temperature of the Ising model on these lattices is exactly calculated for finite iteration steps, and for a range of the scale invariance ratio $n$ from 3 to 1000 and of the number of deleted sites from $(n-2)^2$ to $(n-10)^2$. The critical temperature at the limit of an infinite number of iteration of the segmentation process is asymptotically extrapolated. Comparison is made with results obtained previously by numerical methods. Thermodynamical functions are also calculated and the fractal spectra of the Ising partition functions on several examples of Sierpi\'nski carpets are illustrated.
\end{abstract}


\maketitle

\section{Introduction}

The question of phase transitions on structures of non integer dimension has a long history since the early work of Mandelbrot introducing fractals in Physics. Among the many applications of fractals in most fields of Physics, they have been used as models of structures of non integer dimension for comparing scaling properties to results given by the usual theory of renormalisation in which dimensionality plays a crucial role. Translation invariance is an important hypothesis of the renormalisation approach, not satisfied by structures of non integer dimension which are generally scale invariant rather than translationnally invariant. Then unusual behavior may be expected.

The Sierpi\'nsky Carpet \cite{sier}, and its generalisations to any scale invariance ratio is one of the simplest infinitely ramified fractal structure. An infinite ramification order is a mandatory condition to get a non zero critical temperature to phase transitions \cite{venn, gef1}. An appropriate choice of the scale invariance ratio and of the number and positions of removed sites at each iteration allow to approach any non integer value comprised between $1$ and $2$ for the fractal dimension and a large range of topologies.  And the Ising model is one of the simplest model of phase transition. Within this context, the combination of Sierpi\'nski carpets and the Ising model has been extensively studied to investigate the properties of phases transitions in non integer structures. Calculations involve different methods including real space renormalisation group (RSRG) \cite{gef1, gef2, gef3, bon88, bon89}; high temperature expansions \cite{bon89}; Monte-Carlo simulations either with the Metropolis or with the Wolf or the Swendsen-Wang algorithms combined with a finite size scaling analysis (MCFS) \cite{bhan84, bhan85, angl, bon87, mon98, carm, hsiao00, mon01, pruslois, mon04}; Monte-Carlo simulations combined with renormalisation group (MCRG) \cite{hsiao03}; Monte-Carlo simulations combined with short time dynamic scaling (MCSD) \cite{zheng, bab1, bab2, bab3}. According to references, spins are located either on the center of the sites \cite{bon87, mon98, carm, hsiao00, mon01, pruslois, mon04, bab1, bab2, bab3} or on the vertices of the lattice \cite{venn, gef1, gef2, gef3, bon87, bon88, bon89, bhan84, bhan85, angl, zheng}. Some attempts have been done also on structures extending Sierpi\'nsky carpets in three dimensions: the Menger sponge and its generalizations to any scale invariance ratio \cite{bhan85,hsiao00}. One important goal was to compare the critical properties of the Ising model to those obtained by the analytical continuation of $\varepsilon$-expansions of the renormalisation approach in non integer dimensions \cite{guilzin}: critical exponents and scaling relations. 

One of the difficulty of the Monte-Carlo simulations is to obtain an accurate estimation of the critical temperature, which is necessary to perform the finite size scaling analysis. As fractals are not translationally invariant, the topology changes between two successive iterations (or segmentation steps) of the fractal structure. The mean value of the next nearest neighbours is not strictly the same \cite{mon01, perr}) and consequently the critical temperatures are also different, leading to ``topological scaling corrections'' \cite{mon01}. Moreover, Pruessner and Loison \cite{pruslois} questioned the relevance of finite ssize scaling performed on the successive segmentation steps of a single site, arguing that the size of the stucture is not large enougth to exceed the correlation length in the critical region. These authors developped an alternative approach based on the juxtaposition of several identical networks in the two dimensions of space to increase the size of the elementary cell before applying the usual periodic boundary conditions. For a scale invariance $n=3$ with a single centered removed site, the values of the critical temperatures obtained after this modification are close to those obtained previously, but significantly different according to their respective accuracy, $1.50$ instead of $1.48$. Table \ref{tableref} summarizes the values from litterature of the critical temperature of various Sierpi\'nski carpets, with PBC and spins located on the center of sites, which corresponds to the case treated in this paper. Estimations concentrate close to the same two previous significantly different values, either $1.48$ \cite{mon98, carm, mon01, hsiao03} or $1.50$ \cite{pruslois, bab1, bab3}, leaving a debate opened. 

\begin{table}[tbp]
\setlength{\extrarowheight}{2pt}
\caption{The critical temperatures $T_c$ of various Sierpi\'nsky carpets $SC(n,p)$ with PBC and spins on the center of sites, from litterature in chronological order.}
\label{tableref}
\begin{ruledtabular}
\begin{tabular}{c|lccccc}
$SC$&Authors\hfill&$T_c$&Method&$k_{\rm max}$\\ \tableline
&Bonnier et al.\cite{bon87}&1.54&MCFS&3\\
&Monceau et al. \cite{mon98}&1.482(15)&MCFS&7\\
&Carmona et al. \cite{carm}&1.481(1)&MCFS&7\\
$SC(3,1)$&Monceau et al. \cite{mon01}&1.4795(5)&MCFS&8\\
&Pruessner et al. \cite{pruslois}&1.4992(11)&MCFS&6\\ 
&Hsiao et al. \cite{hsiao03}&1.47946(16)&MCRG&8\\ 
&Bab et al. \cite{bab1}&1.4945(50)&MCSD&6\\
&Bab et al. \cite{bab3}&1.495(5)&MCSD&6\\ \tableline
&Bonnier et al. \cite{bon87}&1.25&MCFS&3\\ 	
&Carmona et al. \cite{carm}&1.077(3)&MCFS&6\\
$SC(4,2)$&Monceau et al. \cite{mon01}&$<$1.049&MCFS&6\\
&Monceau et al. \cite{mon04}&1.13873(8)&MCFS&6	\\
&Bab et al. \cite{bab2}&1.10(1)&MCSD&6\\ 
&Bab et al. \cite{bab3}&1.10(1)&MCSD&6\\ \tableline
&Bonnier et al. \cite{bon87}&2.06&MCFS&3\\
$SC(5,1)$&Monceau et al. \cite{mon01}&2.0660(15)&MCFS&5\\ 
&Bab et al. \cite{bab3}&2.067(2)&MCSD&6\\ \tableline
&Monceau et al. \cite{mon01}&$<$0.808&MCFS&5\\
$SC(5,3)$&Monceau et al. \cite{mon04}&0.96143(11)&MCFS&5\\
&Bab et al. \cite{bab3}&0.83(2)&MCSD&5\\ \tableline
$SC(6,4)$&Bab et al. \cite{bab3}&0.70(5)&MCST&4\\
\end{tabular}
\end{ruledtabular}
\end{table}

To bring some light on this question, we calculate analytically several Sierpi\'nski carpets by a path counting method. We firstly remind the loops counting method on square (lacunary or not) lattices which has been introduced by Kac and Ward \cite{kacward} as an alternative to the algebraic method of Onsager \cite{onsager} to calculate the partition function of the two dimensional Ising on the square lattice. Then this method is applied to Sierpi\'nski carpets with a central hole, but the method is suitable also for other topologies. This leads to the exact calculation of the critical temperatures for values of the segmentation ratio $n$ from $3$ to $1000$ and the number of removed sites $p$ at each step from $1$ to $n-2$. Finally thermodynamical functions of the Ising model on these structures are adressed. Several examples of the fractal spectrum of the Ising partition function on $SC(n,p,k)$ are illustrated.

Notations used in this paper are the following: $SQ(N)$ is the usual square lattice (without empty sites) of size $N\times N$. Sites are numbered according to Fig. \ref{sitenum}. Parameters concerning the square lattice at the limit $N\rightarrow\infty$ will be indexed by $sq$. Generally the lacunary or fractal lattices investigated in this paper are obtained by the juxtaposition of $N^2$ identical patterns in a translation invariant way in both directions of the plan. The reproduced pattern will be called the generating pattern. The number $n$ is devoted to internal characteristics of the generating pattern. $SL(N,n)$ is the lacunary (non random) square lattice of size $N\times N$ with periodic holes of size $(n-1)\times (n-1)$ separated by a single raw of occupied sites in both directions of the plan. Fig. \ref{lac} shows the example of $SL(N,3)$, occupied sites are in grey, the generating pattern is in dark grey. Sites of the generating pattern are numbered as in Fig. \ref{lac}. We have obviously $SQ(N)=SL(N,1)$. Sierpi\'nski carpets of segmentation ratio $n$, with $p^2$ removed central subsites after one segmentation step and after the $k$th iteration of this segmentation process is noted $SC(n,p,k)$. The segmentation step can be omitted when not necessary for understanding: $SC(n,p)$. Fig. \ref{sierpex} (left) illustrates the third segmentation step of $SC(3,1)$. To investigate also a different topology, a more lacunary fractal lattice of segmentation ratio $n=3$, the generating pattern is illustrated on Fig. \ref{sierpex} (right) is also adressed. It is not strictly scale invariant since it is not exactly the reproduction of a same pattern at each segmentation step, its fractal dimension is $\log\left((9+\sqrt{33})/2\right)/\log 3 \approx 1.8184$, it will be noted $SM(3,k)$ (as Sierpin\'ski Modified). Numbering of sites of the generating pattern of Sierpi\'nsky sets is the same as $SQ(n)$, notwithstanding the empty sites, it is illustrated on Fig. \ref{sierpex}. The identity matrix of appropriate dimension is noted $I$ and $d_h$ is the Hausdorff dimension. Following acronyms are used: FBC for free boundary conditions, PBC for periodic boundary conditions.
\vskip1mm
\begin{center}
\includegraphics[width=5cm]{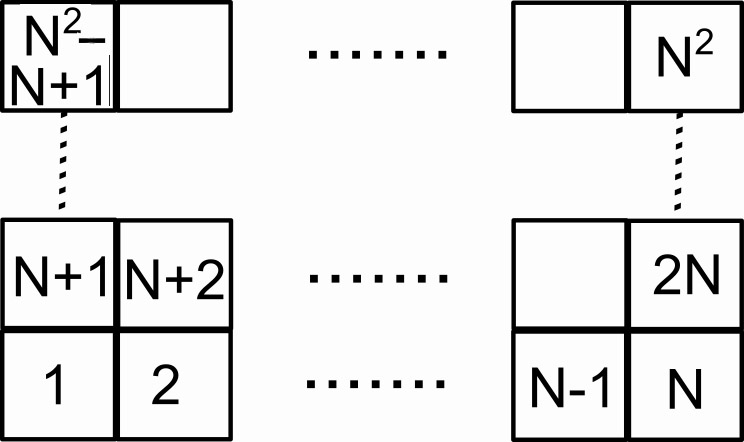}
\captionof{figure}{The numbering of sites of a generating pattern of Sierpin\'ski sets with a segmentation ratio $n$.} 
\label{sitenum}
\end{center}

\begin{center}
\includegraphics[width=6cm]{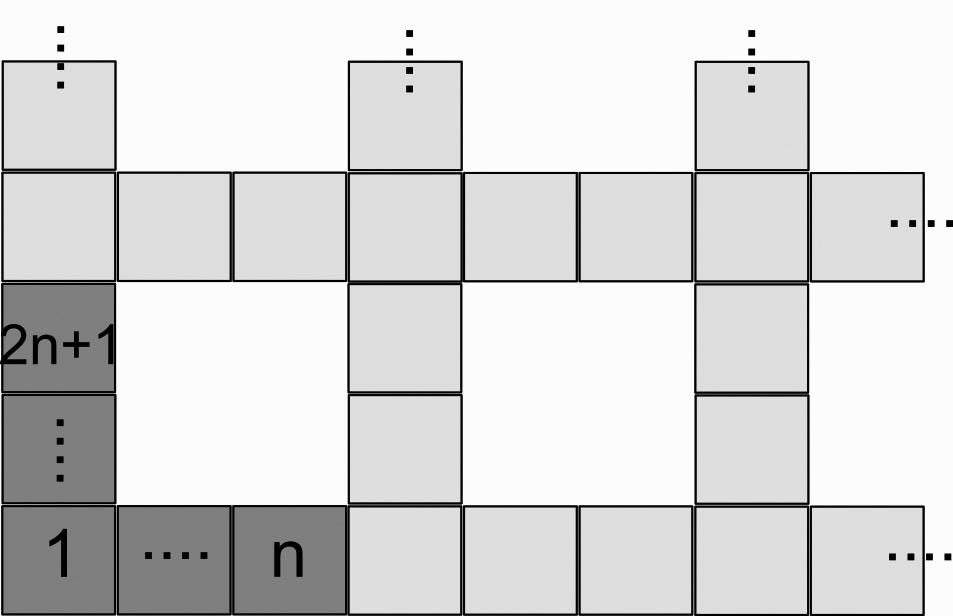}
\captionof{figure}{Illustration of $SL(N,n)$ for $n=3$, occupied sites are in grey, the generating pattern is in dark grey.} 
\label{lac}
\end{center}

\begin{center}
\centerline{\includegraphics[width=4.12cm]{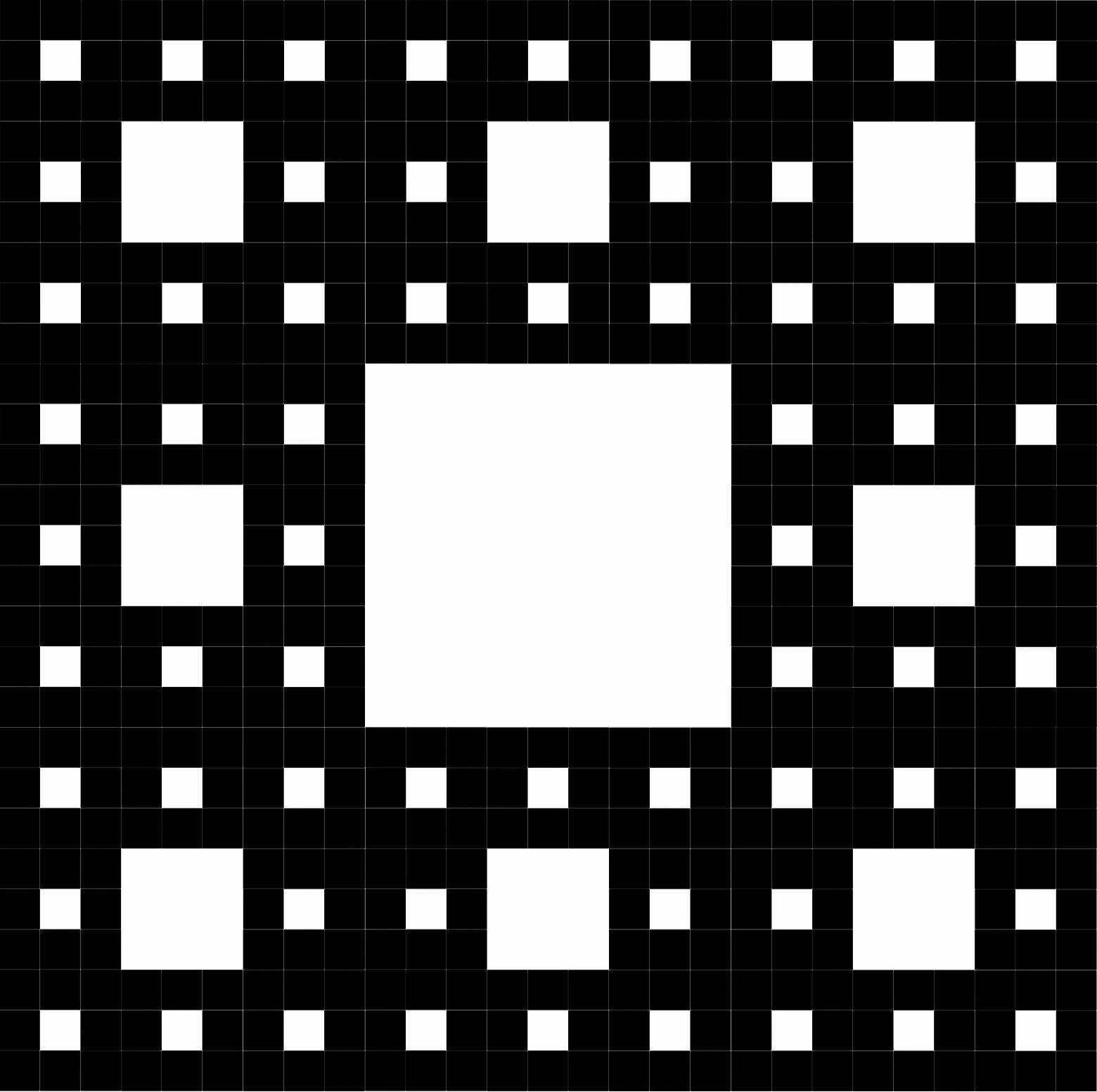}\hfill \includegraphics[width=4.25cm]{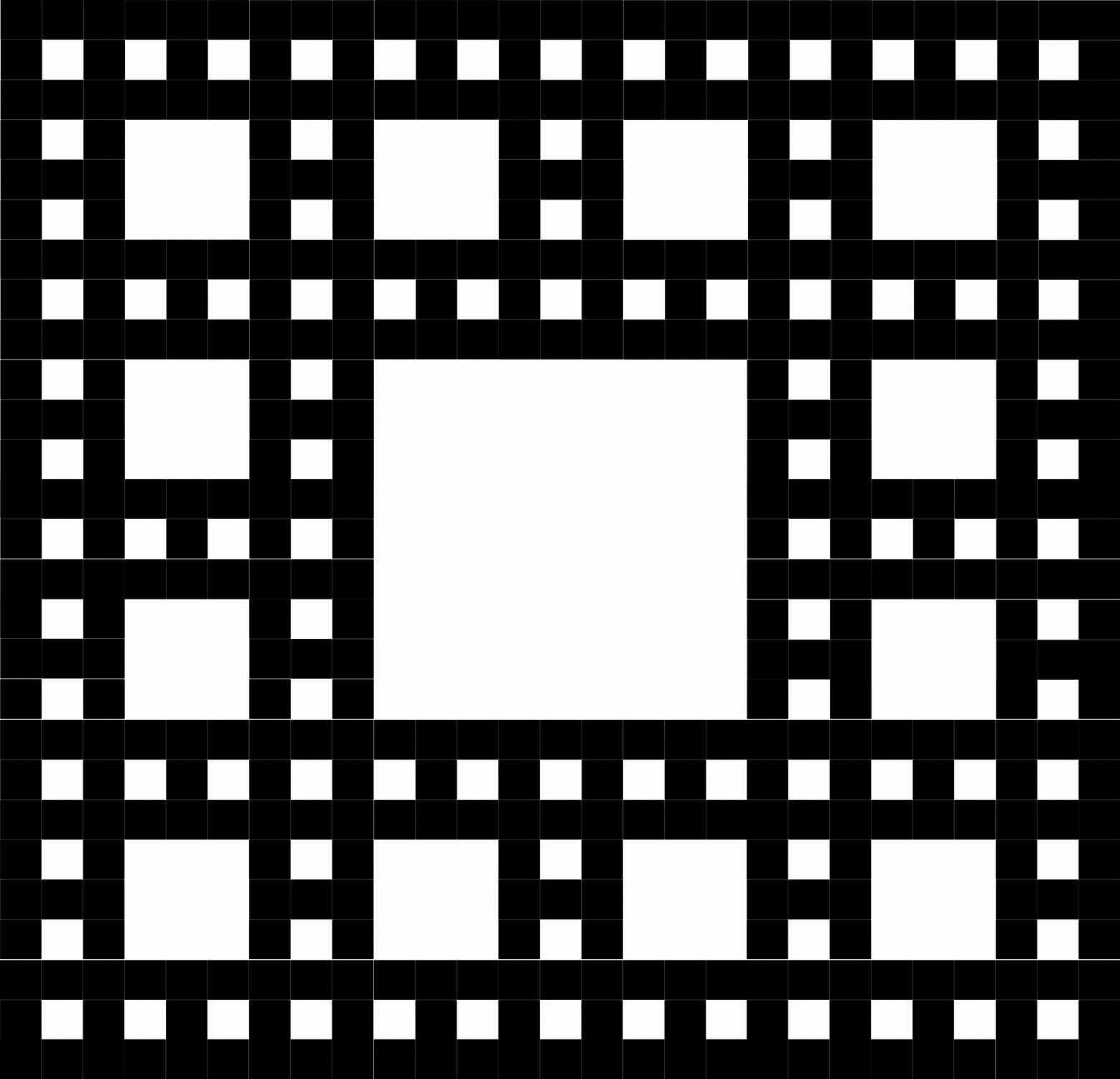}}
\captionof{figure}{The third iteration steps $SC(3,1,3)$ (left) and $SM(3,3)$ (right).} 
\label{sierpex}
\end{center}

\section{Partition function of the Ising model and closed pathes counting}

Generally, the partition function of the Ising model on any network is \cite{binney}:
$$Z=\sum_{\{s_i\}}\prod_{i>j}e^{-\beta Js_is_j}$$
where the sum runs over all spin configurations, the product over all pairs of nearest neighbours spins, $J$ characterizes the strength of the interaction between neighbouring spins. It is a equivalent to the problem of counting closed pathes (loops) on the network \cite{kacward}:
\begin{equation}
Z=2^{\cal N}\cosh\left({1\over T}\right)^{{\cal N}_z}\sum_{l=0}^\infty g(l)v^l
\label{partising}
\end{equation}
where $\cal N$ is the number of spins, ${\cal N}_z$ is the number of nearest neighbours pairs, $T$ is the reduced temperature: $1/\beta J$, $v=\tanh(\beta J)=\tanh(1/T)$ and $g(l)$ is the number of loops of length $l$ (counted as a multiple of the distance between two neighbouring sites). In the next developments, we will call ${\cal P}(v)=\sum_{l=0}^\infty g(l)v^l$ the {\sl partition polymonial}.

Kac and Ward \cite{kacward} implemented this method to $SQ(N)$ in the following way (a variant using pfaffians has also been proposed \cite{hurst, coywu}, and the method has been subsequently simplified\cite{binney, glasser}). The basic feature is to attribute to each site a coefficient which connects the direction from where the path enter the site to the direction it goes out in order that the product of coeffients of all sites of a closed, non self-intersecting path gives $-1$. The appropriate set of coefficients are: $1$ if the path continues the same direction, $e^{i\pi\over 4}$ for a left turn, $e^{-i\pi\over 4}$ for a right turn, $0$ for the backward direction. On a finite lattice with FBC, this procedure ensures that, when performing the product over sites of all pathes and the sum over all possible pathes, each closed non self intersecting loop is counted the number of times it should be \cite{sherm, burg}. To implement this counting method on the whole lattice, four $4\times 4$ matrices are introduced linking each site to its nearest neighbours. They contain the appropriate coefficients placed according to the directions ``in'' and ``out'' and they are indexed by the direction of the site from where the path is coming from, ordered as $-x$, $-y$, $y$, $x$:
\begin{equation}
\begin{small}
\begin{aligned}
\let\quad\thinspace
M_{-x}&=\begin{pmatrix}0&0&0&0\\ 0&0&0&0\\ 0&0&0&0\\ 0&e^{i\pi\over 4}&e^{-{i\pi\over 4}}&1\end{pmatrix}&
M_{-y}&=\begin{pmatrix}0&0&0&0\\ 0&0&0&0\\ e^{-{i\pi\over 4}}&0&1&e^{i\pi\over 4}\\ 0&0&0&0\end{pmatrix}\\
M_y&=\begin{pmatrix}0&0&0&0\\ e^{i\pi\over 4}&1&0&e^{-{i\pi\over 4}}\\ 0&0&0&0\\ 0&0&0&0\end{pmatrix}&
M_x&=\begin{pmatrix}1&e^{-{i\pi\over 4}}&e^{i\pi\over 4}&0\\ 0&0&0&0\\ 0&0&0&0\\ 0&0&0&0\end{pmatrix}\\
\end{aligned}
\end{small}
\end{equation}

These matrices are themselves placed in a matrix $\cal M$ of all sites and the partition polynomial is 
$${\cal P}_N(v)=\sqrt{\left| I-v{\cal M}\right|}$$

The spectrum of the partition function (including the root corresponding to the critical temperature) is given by the set of roots of this polynomial in the variable $v$.

This process may be extended to any planar lacunary square lattices, setting to $0$ the $4\times 4$ matrices corresponding to empty sites. We calculate here the partition function of $SL(N,n)$. The generating pattern has $2n-1$ occupied sites (Fig. \ref{lac}), the dimension of the matrix ${\cal M}_{N,n}$ is $4N(2n-1)$. PBC make ${\cal M}_{N,n}$ cyclic which allows to calculate $|I-v{\cal M}_{N,n}|$ as the product of $N^2$ $4(2n-1)\times 4(2n-1)$ determinants according to a process which has been explained in many references for $n=1$ \cite{kacward,binney, glasser}. Let us note 
\begin{equation*}
\begin{array}{llllll}
M'_{-x}&=&e^{i p\pi\over N}M_{-x}, &\qquad M'_{-y}&=&e^{i q\pi\over N}M_{-y},\\
M'_y&=&e^{-{i q\pi\over N}}M_y, &\qquad M'_x&=&e^{-{i p\pi\over N}}M_x,
\end{array}
\end{equation*}
the corresponding determinant is the product over $p$ and $q$ varying from $1$ to $N$ of determinants involving the matrices:
\begin{equation*}
\begin{small}
\let\quad\thinspace
\bordermatrix{&\sss 1&\sss 2&\sss 3&\dots&\sss n-1&\sss n&\sss n+1&\sss n+2&\dots&\sss 2n-2&\sss 2n-1\cr
\hfill\sss  1\hfill&0&M_{-x}&0&\dots&0&M'_x&M_{-y}&0& \dots&0&M'_y\cr
\hfill\sss  2\hfill &M_x&0&M_{-x}&0&\dots&\dots&\dots&\dots&\dots&\dots&0\cr
\hfill\sss  3\hfill&0&M_x&0&\ddots&\ddots&&&&&&\vdots\cr
\hfill \vdots\hfill &\vdots&0&\ddots&\ddots&\ddots&\ddots&&&&&\vdots\cr
\hfill\sss  n-1\hfill &0&\vdots&\ddots\ &\ddots\ \ &0&M_{-x}&\ddots&&&&\vdots\cr
\hfill\sss  n\hfill &M'_{-x}&\vdots&&\ddots\ &M_x&0&0&\ddots&&&\vdots\cr
\hfill\sss  n+1\hfill &M_y&\vdots&&&\ \ddots&0&0&M_{-y}&\ \ddots&&\vdots\cr
\hfill\sss  n+2\hfill &0&\vdots&&&&\ \ddots&M_y&\ddots&\ddots&\ \ddots&\vdots\cr
\hfill \vdots\hfill &\vdots&\vdots&&&&&\ \ddots&\ddots&\ddots&\ddots&0\cr
\hfill\sss  2n-2\hfill &0&\vdots&&&&&&\ \ddots&\ddots&0&M_{-y}\cr
\sss 2n-1\hfill &M'_{-y}&0&\dots&\dots&\dots&\dots&\dots&\dots&0&M_y&0\cr}
\end{small}
\end{equation*}

Given that their eigenvectors of eigenvalue $\lambda$ are $V=(V_i)_{1\leq i\leq 2n-1}$ where the $(2n-1)$ 4-vectors $V_i$ (one by site of the generating pattern) have the following structure,
\begin{equation*}
\begin{small}
\begin{array}{ccc}
&{\rm for}\ 1<i\leq n&{\rm for}\ n<i\leq 2n-1\\ \\
V_1=\begin{pmatrix}a\\b\\c\\d\end{pmatrix}&{V_i}=\begin{pmatrix}a\lambda^{(n-i+1)}e^{2 i \pi p\over N}\\0\\0\\ d\lambda^{(i-1)}\end{pmatrix}& {V_i}=\begin{pmatrix}0\\b\lambda^{(2n-i)}e^{2 i \pi q\over N}\\c\lambda^{(i-n)}\\0\end{pmatrix},
\end{array}
\end{small}
\end{equation*}
and noting
\begin{equation}
M(N,p,q)=M'_{-x}+M'_{-y}+M'_y+M'_x
\label{matred}
\end{equation}
the final product is 
\begin{equation}
\label{pseudoI2d}
\begin{small}
\begin{aligned}
P_{N,n}(v)&=\bigg[\prod_{p,q=1}^N\left|I-v^nM (N,p,q)\right|\bigg]^{1\over 2}\\
&=\bigg[\prod_{p,q=1}^N(1+v^{2n})^2-2v^n(1-v^{2n})f(N,p,q)\bigg]^{1\over 2}\\
\end{aligned}
\end{small}
\end{equation}
with $f(N,p,q)=\cos{2p\pi\over N}+\cos{2q\pi\over N}$.

It should be noticed that $P_{N,n}(v)=P_{N,1}(v^n)$. The lacunary square lattice is equivalent to a filled square lattice in which the parameter $v$ is replaced by $v^n$. That results from the structure of the generating pattern of the lacunary square lattice which funnels pathes going through in a single direction, either $Ox$ or $Oy$. Changes of direction can occur only on the first site of the pattern, like if there was a single site. Compared to a single site pattern, only the length of pathes changes, not the topology. This equivalence will be exploited later for the calculation of the magnetization.

However, $P_{N,n}(v)$ is not exactly the partition polynomial ${\cal P}_{N,n}(v)$ of the Ising model on $SL(N,n)$. The process of indexing loops is rigorously demonstrated for FBC while PBC are necessary to calculate the determinant. PBC introduce loops looping the torus (Fig. \ref{torus}) for which the product of all sites coefficients is $1$ and not $-1$ as it should be. 
\vskip1mm
\begin{center}
\includegraphics[width=4.5cm, height=2cm]{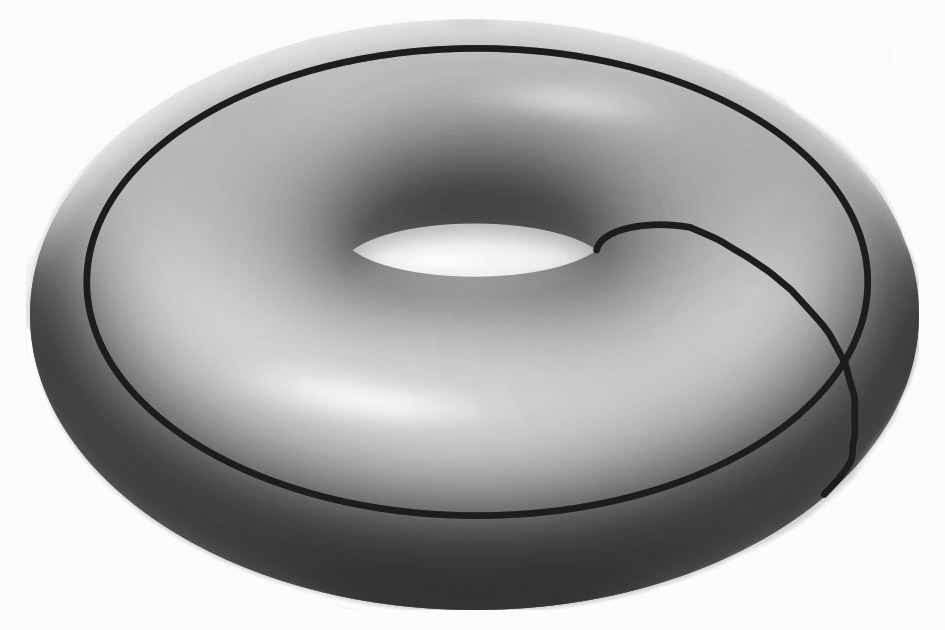}
\captionof{figure}{Two pathes looping the torus which are wrongly counted in PBC.} 
\label{torus}
\end{center}

These boundary effects vanish at the limit ($N\rightarrow\infty$) giving the exact partition function of the infinite lattice, but they cannot be ignored when finite size lattices are concerned. For this reason, we will call $P_{n,N}(v)$ the {\sl pseudopartition polynomial}. Additional terms necessary to obtain the exact partition polynomial ${\cal P}_{n,N}$ of the Ising model on $SQ(N)$ with PBC have been subsequently calculated \cite{pottsw}, however they will not play a role in the next developments of this paper.

A consequence is that the spectrum of the pseudopartition polynomial of $SL(N,n)$ is immediately deduced from the spectrum of the pseudopartition polynomial of $SQ(N)$. Let us call $\theta$ the function
\begin{equation}
\begin{small}
\theta_\pm(N,p,q)={\rm Atan}\left({-\sqrt{4-f^2}\pm\sqrt{\vert f\vert(2-\vert f\vert)}\over 2-\vert f\vert\pm\sqrt{\vert f\vert(2+\vert f\vert)}}\right),
\end{small}
\end{equation}
then the spectrum $v^n$ is given by the set of the following four values: 
\begin{equation}
\begin{small}
\begin{aligned}
v_+^n&=-{\rm sgn}(f(N,p,q))+\sqrt{2}\,e^{i\theta_+(N,p,q)}&\kern-2mm\\
v_{-}^n&=-{\rm sgn}(f(N,p,q))+\sqrt{2}\,e^{i\theta_{-}(N,p,q)}&\kern-2mm=-{1\over v_+^n}\\ 
{v_+^n}^*&=-{\rm sgn}(f(N,p,q))+\sqrt{2}\,e^{-i\theta_+(N,p,q)}&\kern-2mm\\
{v_{-}^n}^*&=-{\rm sgn}(f(N,p,q))+\sqrt{2}\,e^{-i\theta_{-}(N,p,q)}&\kern-2mm=-{1\over {v_+^n}^*}\\
\end{aligned}
\end{small}
\end{equation}

For any value of $N$, all roots $v^n$ are located on two circles in the complex plan, centered respectively on $-1$ and $1$ (according to the sign of $f(N,p,q)$) and of radius $\sqrt{2}$. Fig. \ref{isingplot} shows these two circles and plots of the spectrum of the pseudopartition polynomial $P_5(v)$ (squares) and of the exact partition polynomial ${\cal P}_5(v)$ (triangles) of $SQ(5)$. The spectrum of ${\cal P}_N(v)$ converges toward the two circles when $N\rightarrow\infty$.

\vskip1mm
\begin{center}
\includegraphics[width=7cm]{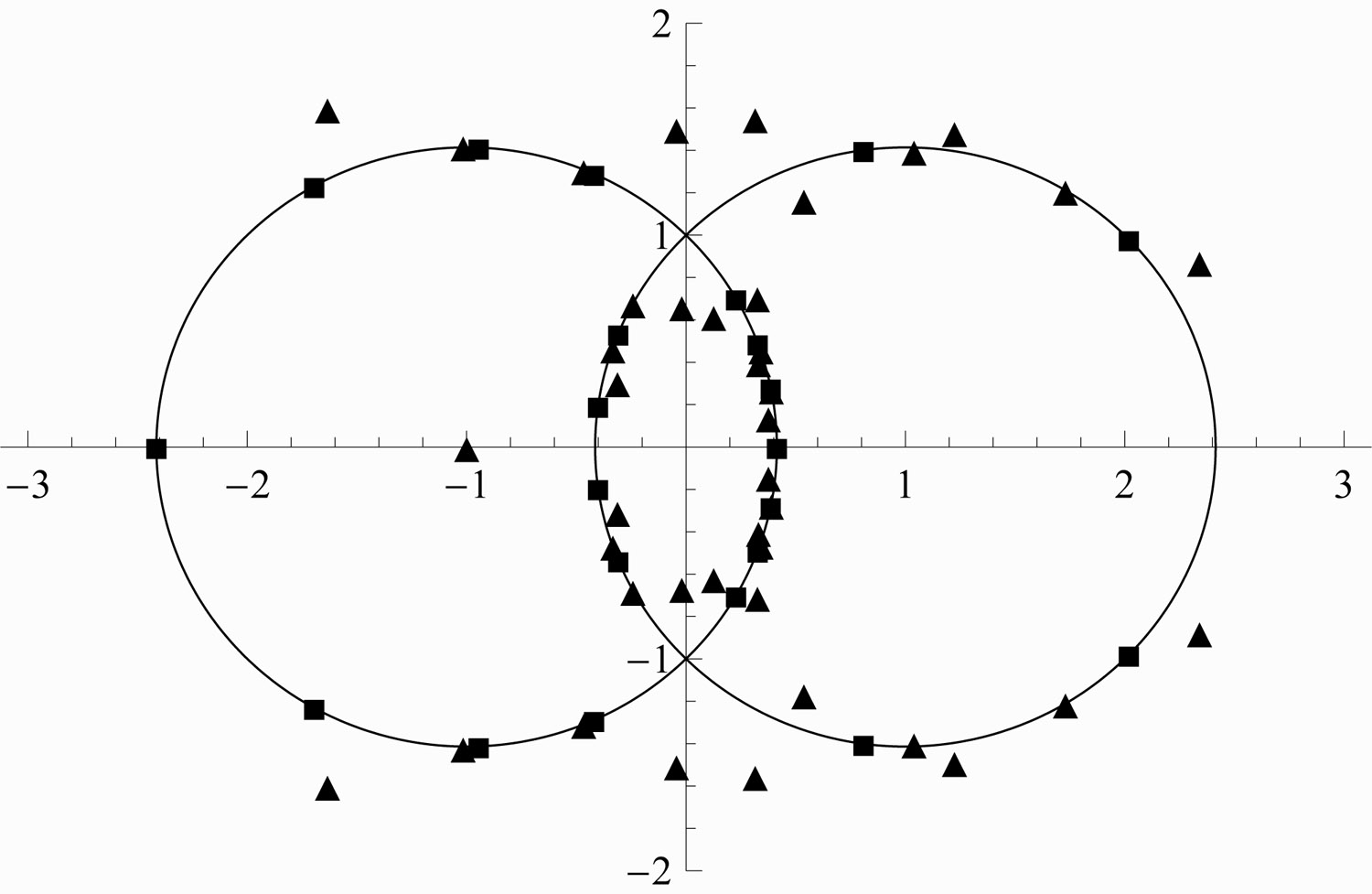}
\captionof{figure}{The combined plots in the complex plan of the roots of $P_5(v)$ (squares), ${\cal P}_5(v)$ (triangles) and the two circles which contains the roots of $P_N(v)$, for all $N$.} 
\label{isingplot}
\end{center}
\vskip1mm

Fig. \ref{pseudopartition} shows the plot of $P_{n,N}(v)$ versus $v^n$ for $N=1$ to $N=7$. 

\vskip1mm
\begin{center}
\includegraphics[width=8.5cm]{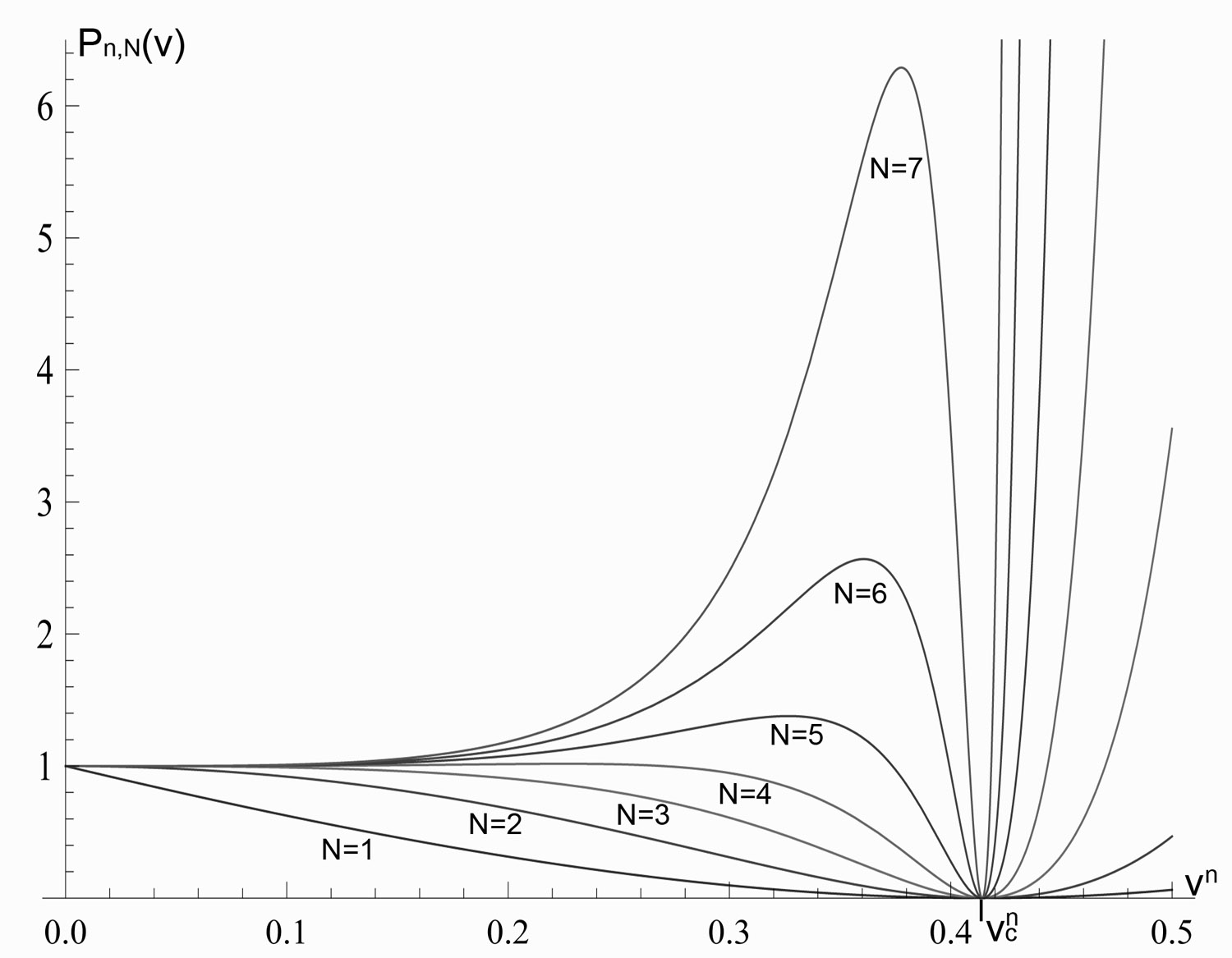}
\captionof{figure}{The pseudopartition polynomial $P_{N,n}(v)$ of the Ising model on $SL(N,n)$, versus $v^n$ for $1\leq N\leq 7$.} 
\label{pseudopartition}
\end{center}
\vskip1mm

For all values of $N$, $P_{N,n}(v)$ vanishes for the same value of $v^n$. Indeed, all $P_{N,n}(v)$, share the common factor $P_{1,n}(v)$, the partition polynomial of a ``single pattern lattice'' (a ``single site'' lattice when $n=1$):
\begin{equation}
P_{1,n}(v)=1-2v^n-v^{2n}
\label{pseudn1} 
\end{equation}
whose positive real root is $v_c=(\sqrt{2}-1)^{1/n}$, giving the critical temperature at the liimit thermodynamical limit (this root is the same for all $N$, then also for the limit $N\rightarrow\infty$):
\begin{equation}
T_c={1\over\hbox{Atanh}(v_c)}={2\over \log\left({1+(\sqrt{2}-1)^{1\over n}\over 1-(\sqrt{2}-1)^{1\over n}}\right)}
\label{tcrit}
\end{equation}

For $n=1$, ${v_c}_{sq}=\sqrt{2}-1$ gives the well known value: 
$${T_c}_{sq}={2\over \log\left(1+\sqrt{2}\right)}\approx 2.2692$$

Then a fast and easy way to calculate the critical temperature of any square lattice (lacunary or not) at the limit $N\rightarrow\infty$ is to find the real positive root of the pseudopartition polynomial of a single pattern lattice. The pseudopartition polynomial can be considered as a partition polynomial renormalized of the boundaries effects. This property will be used to calculate the critical temperature of the Sierpi\'nski carpets, which is the purpose of the next section.

\section{Partition function of the Ising model on $SC(n,p,k)$}

Investigated lattices are composed by the juxtaposition of $N^2$ generating pattern $SC(n,p)$ in the two directions of the plane, with PBC. The generating pattern has $(n^2-p^2)^k$ ocupied sites. The matrix ${\cal M}_{SC(n,p,k)}$ is obtained from the matrix ${\cal M}_{(nN)^{2k}}$ of the corresponding square lattice by setting to $0$ the $4\times 4$ matrices of each removed site. Lines and columns corresponding to these empty sites can be omitted, which reduces the dimension of the matrix to $4N^2(n^2-p^2)^k$. As for the square lattice, the determinant may be factorized by translation invariance in a product of $N^2$ determinants of dimension $4(n^2-p^2)^k$. Detailed calculations are performed for $SC(3,1,1)$, other values of $n$, $p$ and $k$ are treated in a similar way. 

\subsection{The pseudopartition polynomial of $SC(3,1,1)$}

In $SC(3,1, 1)$ a single site is removed. Once reduced by translation invariance, the pseudopartition polynomial is the square root of the product of the $N^2$ $32N^2 \times 32N^2$ determinants of the following matrices. The site number five, which is the single empty site is omitted.
\begin{equation}
\let\quad\thinspace
\begin{array}{l}
M_{N,SC(3,1,1)}(p,q)=\\
\bordermatrix{&\sss 1&\sss 2&\sss 3&\sss 4&\sss 6&\sss 7&\sss 8&\sss 9\cr
\sss 1&0&M_{-x}&M'_x&M_{-y}&0&M'_y&0&0\cr
\sss 2&M_x&0&M_{-x}&0&0&0&M'_y&0\cr
\sss 3&M'_{-x}&M_x&0&0&M_{-y}&0&0&M'_y\cr
\sss 4&M_y&0&0&0&M'_x&M_{-y}&0&0\cr
\sss 6&0&0&M_y&M'_{-x}&0&0&0&M_{-y}\cr
\sss 7&M'_{-y}&0&0&M_y&0&0&M_{-x}&M'_x\cr
\sss 8&0&M'_{-y}&0&0&0&M_x&0&M_{-x}\cr
\sss 9&0&0&M'_{-y}&0&M_y&M'_{-x}&M_x&0\cr}
\end{array}
\label{fullmatrix}
\end{equation}

Leading to the pseudopartition polynomial
\begin{equation}
\begin{small}
\begin{aligned}
P_{N,SC(3,1,1)}(v)={\displaystyle\prod_{p,q=1}^N}\left|I-v.M_{N,SC(3,1)}(p,q)\right|^{1\over 2}\qquad\\
={\displaystyle\prod_{p,q=1}^N}\biggl[Q_1(v)+Q_2(v)\left(\cos\left({2\pi p\over N}\right)\right.\ \\
\left.+\cos\left({2\pi q\over N}\right)\right)+Q_3(v)\cos\left({2\pi p\over N}\right)\cos\left({2\pi q\over N}\right)\\
\qquad+Q_4(v)\left(\cos\left({2\pi p\over N}\right)^2+\cos\left({2\pi q\over N}\right)^2\right)\biggl]^{1\over 2}
\end{aligned}
\end{small}
\label{pseudosier}
\end{equation}
where
\begin{equation*}
\begin{small}
\begin{array}{lll}
Q_1(v)=&(1+v^2)^2(1-2v^2+13 v^4-8 v^6+126v^8+108v^{10}\\
&\qquad\qquad+474v^{12}+248v^{14}+57v^{16}+6v^{18}+v^{20})\\ \noalign{\vglue3pt}
Q_2(v)=&\ -4v^3(1-v^2)^2(1+3v^2)(1+v^2+2v^4)\\
&\qquad\qquad\qquad\qquad(1+2v^2+8v^4+4v^6+v^8)\\ \noalign{\vglue3pt} 
Q_3(v)=&-4v^6(1-v^2)^4(1+v^2)(7+11v^2+13v^4+v^6)\\ \noalign{\vglue3pt}
Q_4(v)=& +4v^6(1-v^2)^5(1+v^2)(1+3v^2)\\
\end{array}
\end{small}
\end{equation*}

The critical temperature is given by the real positive root of
\begin{equation}
\label{partSC31}
\begin{aligned}
P_{1,SC(3,1,1)}(v)\ =&\ 1-4v^3+5v^4-16v^5-10v^6\\&\ -20v^7+v^8-24v^9+2v^{10}+v^{12}\\
\end{aligned}
\end{equation}
which is $v_c\approx 0.4960$ giving $T_c=1/\hbox{Atanh}(v_c)\approx 1.8384$.

For small values of $n$ and/or small values of $k$, $P_{1,SC(n,p,k)}$ may be calculated explicitely and roots are calculated numerically. For higher values of $n$ or $k$, the spectrum may be deduced directly from the numerical evaluation of the eigenvalues of the exact matrix. But for very large values of $n$ or $k$, the dimension of the matrix increases quickly. To get access  to the spectrum of the partition function, the dimension of the matrix should be reduced and this can be achieved in two ways. First exploiting the square symmetry of the generating pattern, that is its invariance under the group $D4$ (especially the rotation of angle $\pi/4$). Second, by removing the zero eigenvalues which come from dangling bonds of sites contiguous to empty sites. We will investigate successively these two ways.

\subsection{Rotation invariance}

The appropriate four dimensional representation of a rotation of $\pi/4$ is generated by a $4\times 4$ matrix of the form 
\begin{equation}
\sigma=\alpha\begin{pmatrix}0&0&1&0\\ 1&0&0&0\\ 0&0&0&1\\ 0&1&0&0\\\end{pmatrix}
\label{sigma}
\end{equation}
which achieves a circular permutation of the four directions in the plan, with the constraint that $\sigma^4=I$ and the condition $\alpha^4=1$. There are four possible values of $\alpha$: $\alpha_1=1$, $\alpha_2=-1$, $\alpha_3=i$, $\alpha_4=-i$ and four corresponding matrices $\sigma_i$, ($1\leq i\leq 4$) according to equation \eqref{sigma}. 

The eigenvectors of the matrix $M_{1,SC(n,p,k)}$ are composed of $(n^2-p^2)^k$ 4-vectors numbered according to sites $V_1, V_2......V_{n^2}$ (with the exclusion of empty sites) which can be associated four by four as the images of an initial eigenvector by the successive rotations $\sigma_i$, $\sigma_i^2$, $\sigma_i^3$. For $SC(3,1,1)$, this gives (Fig. \ref{eigen}): 
$$V_i=(V_1, V_2, \sigma_i.V_1, \sigma_i^3.V_2, \sigma_i.V_2, \sigma_i^3.V_1, \sigma_i^2.V_2, \sigma_i^2.V_1)$$. 
\begin{center}
\includegraphics[width=4cm]{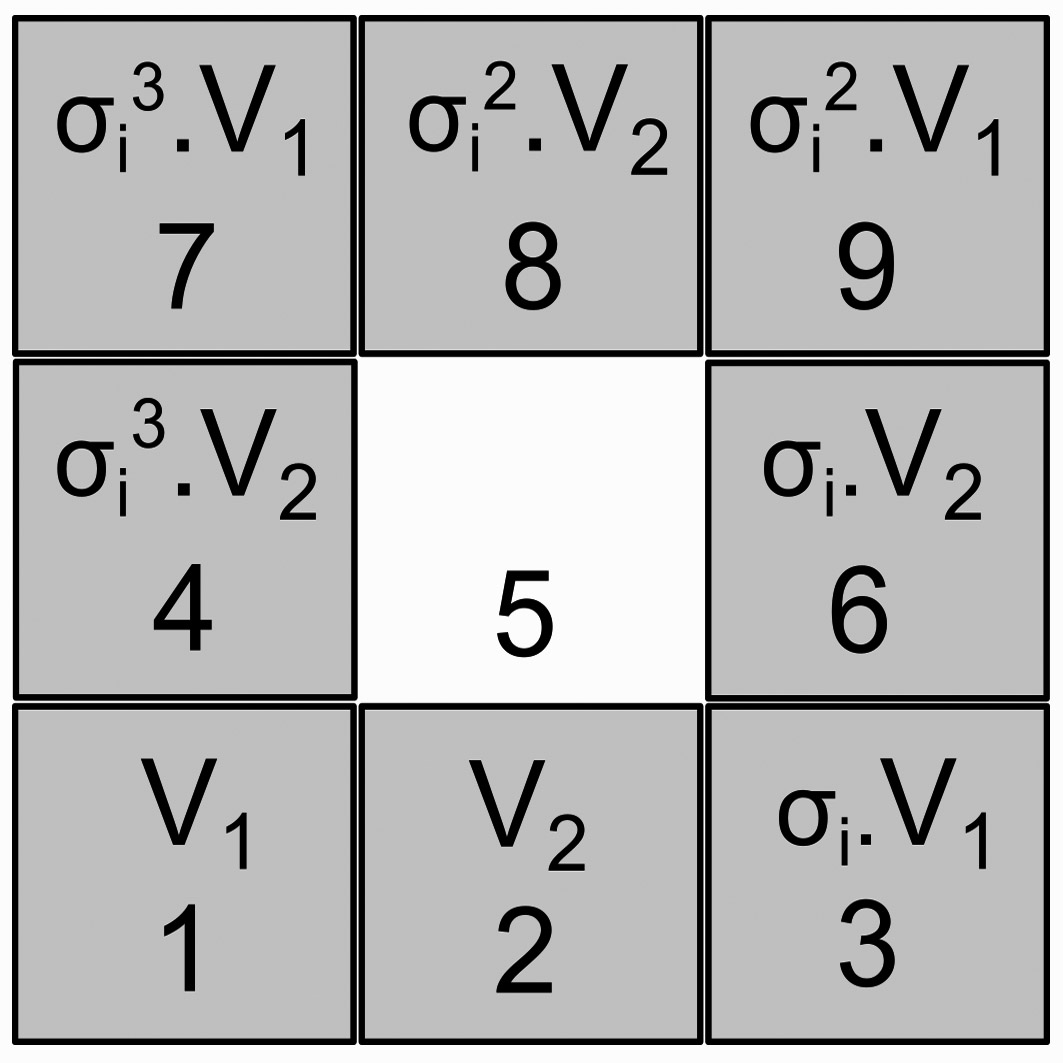}
\captionof{figure}{The structure $SC(3,1,1)$ with, on each site, the site number (down) and the corresponding part of the $i$th eigenvector of $M_{SC(3,1,1)}$ (up).} 
\label{eigen}
\end{center}
\vskip1mm

Then, for $SC(n,p,k)$ (a single pattern lattice) with a total number of non empty sites $(n^2-p^2)^k$, the $4(n^2-p^2)^k\times 4(n^2-p^2)^k$ determinant is transformed in the product of four $(n^2-p^2)^k\times (n^2-p^2)^k$ determinants. For $SC(3,1,1)$, we obtain the four matrices
\begin{equation*}
M_{SC(3,1,1),i}(v)=\begin{pmatrix}M_x.\sigma_i+M_y.\sigma_i^3& M_{-x}+M_{-y}.\sigma_i^3\\
M_x+M_{-x}.\sigma_i&M_y.\sigma_i^2\\ \end{pmatrix}
\end{equation*}

The pseudopartition polynomial of $SC(3,1,1)$ is:
$$P_{1,SC(3,1,1)}(v)=\sqrt{\prod_{i=1}^4\left|I-vM_{SC(3,1,1),i}\right|}$$ 

Polynomials $\left|I-vM_{SC(3,1,1),i}\right|$ are equal two by two: 
$$\displaylines{\left|I-vM_{SC(3,1,1),1}\right| = \left|I-vM_{SC(3,1,1),3}\right|= Q^{-}_{SC(3,1,1)}(v)\cr
\left|I-vM_{SC(3,1,1),2}\right|= \left|I-vM_{SC(3,1,1),4}\right|= Q^+_{SC(3,1,1)}(v)\cr}$$ 
with
\begin{equation*}
\begin{array}{lr}
Q^\pm_{SC(3,1,1)}(v) = &1\pm \sqrt{2}v +(1\pm \sqrt{2})v^2\pm \sqrt{2}v^3\\&+(5\pm 3\sqrt{2})v^4 \pm 2\sqrt{2}v^5+v^6\\
\end{array}
\end{equation*}
so that equation \eqref{partSC31} can be written
$$P_{1, {SC(3,1,1)}}(v) = Q^+_{SC(3,1,1)}(v).Q^{-}_{SC(3,1,1)}(v)$$

For all investigated lattices $SC(n,p,k)$, a single real root of the pseudopartition polynomial occurs between $0$ and $1$, which avoid any ambiguity to identify the critical temperature among all other roots, and it is always a root of $Q^-$, but we have no general demonstration.

\subsection{Dangling bonds}

To each site contiguous to an empty site, the direction of a path going to the empty site gives a zero eigenvalue corresponding to an eigenvector with all components equal to zero except the four components located on the position of the corresponding site. According to the position of the empty site relatively to the position of the considered site: $-x$, $-y$, $y$, $x$, the four non zero components of the eigenvectors are $V_{-x}$, $V_{-y}$, $V_y$, $V_x$, respectively and according to Fig. \ref{eigenpos}:

\begin{equation*}
\begin{pmatrix}0\\e^{-{i\pi\over 4}}\\e^{i\pi\over 4}\\-1\end{pmatrix},\ \ \quad\begin{pmatrix}e^{i\pi\over 4}\\0\\-1\\e^{-{i\pi\over 4}}\end{pmatrix},\ \ \quad\begin{pmatrix}e^{-{i\pi\over 4}}\\-1\\0\\e^{i\pi\over 4}\end{pmatrix},\ \ \quad\begin{pmatrix}-1\\ e^{i\pi\over 4}\\e^{-{i\pi\over 4}}\\0\end{pmatrix}
\end{equation*}

\begin{center}
\includegraphics[width=8.5cm]{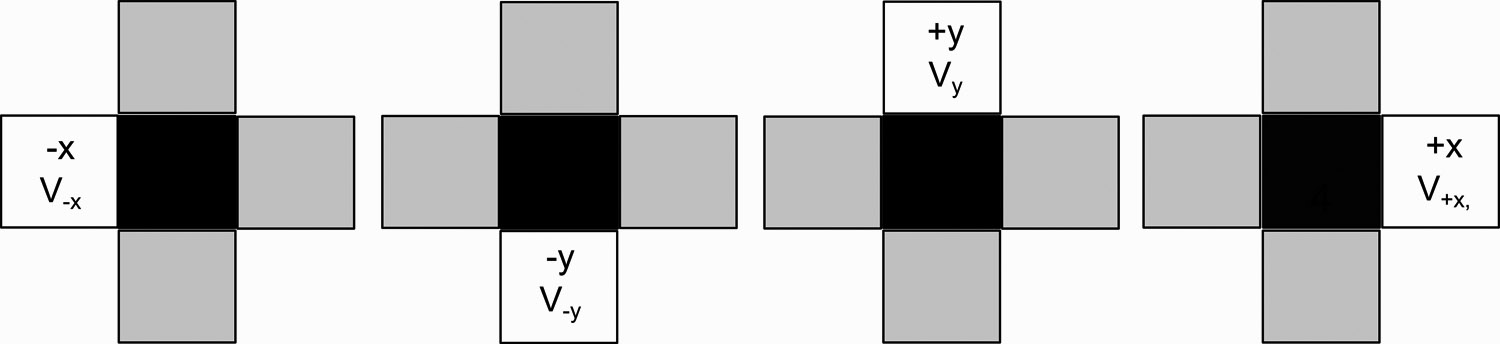}
\captionof{figure}{The four possible configurations of a site contiguous to an empty site with the corresponding eigenvector of zero eigenvalue. In black: the considered site; in grey: occupied sites; in white: the empty site.} 
\label{eigenpos}
\end{center}

Factorizing the part of the matrix corresponding to these eigenvectors reduces the dimension of the determinants (already reduced by rotation invariance) by two times the number of dangling bonds, that is from $(n^2-p^2)^k$  to
\begin{equation*}
\begin{small}
(n^2-p^2)^k-2p{(n^2-p^2)^k-n^k\over n^2-p^2-n}+2p^2{(n^2-p^2)^{k-1}-n^{k-1}\over n^2-p^2-n}
\end{small}
\end{equation*}
for $n-p=2$, or
\begin{equation*}
\begin{small}
(n^2-p^2)^k-2p{(n^2-p^2)^k-n^k\over n^2-p^2-n}
\end{small}
\end{equation*}
for other values of $n-p$.

\section{The critical temperature of Sierpi\'nski carpets}

Table \ref{tab:tabletemp} contains the critical temperature of the Ising model on $SC(n,p,k)$ for a wide range of values of $n$, $p$,~$k$. 

\begin{table}[tbp]
\caption{\label{tab:tabletemp}The critical temperatures of the Sierpi\'nski carpet $SC(n,p,k)$ for $3\leq n\leq 1000$ and $1\leq p\leq n-2$.}
\begin{ruledtabular}
{\setlength{\jot}{4pt}\setlength{\tabcolsep}{0pt}
\begin{tabular}{c|cccccccc}
\multirow{2}{1mm}{$p$}&\multirow{2}{1mm}{$n$}&\multirow{2}{2mm}{$d_h$}&\multicolumn{6}{|c}{$T_c$}\\ \cline{4-9}
&&&\multicolumn{1}{|c}{$k=1$}&$k=2$&$k=3$&$k=4$&$k=5$&$k=\infty$\\ \tableline
\multirow{23}{2mm}{\rotatebox{90}{$n-2$\;}}&3&1.8928&1.8384&1.6538&1.5676&1.5256&1.5045&1.478\\
&4&1.7925&1.6213&1.3689&1.2502&1.1845&&1.113\\
&5&1.7227&1.4875&1.19857&1.0579&0.9882&&0.876\\
&6&1.6720&1.3946&1.0832&0.9309&&&0.745\\
&7&1.6332&1.3252&0.9997&0.8431&&&0.645\\
&8&1.60251&1.2709&0.9363&0.7793&&&0.588\\
&9&1.5773&1.2267&0.8825&0.7308&&&0.546\\
&10&1.5563&1.1899&0.8461&0.6926&&&0.484\\
&11&1.5384&1.1585&0.8126&0.6621&&&0.447\\
&12&1.5229&1.1314&0.7846&&&&\\
&13&1.5093&1.1076&0.7599&&&&\\
&14&1.4972&1.0864&0.7387&&&&\\
&15&1.4864&1.0675&0.7197&&&&\\
&20&1.4456&0.9954&0.6515&&&&\\
&25&1.4180&0.9460&0.6069&&&&\\
&30&1.3976&0.9091&0.5748&&&&\\
&50&1.3492&0.8195&0.5007&&&&\\
&70&1.3229&0.7693&&&&\\
&100&1.2988&0.7222&&&&\\
&200&1.2607&0.6449&&&&\\
&500&1.2227&0.5642&&&&\\
&1000&1.2005&0.5150&&&&\\ \tableline
\multirow{13}{2mm}{\rotatebox{90}{$n-4$\;}}&5&1.9746&2.1193&2.0790&2.0690&&&2.065\\
&6&1.9343&2.0097&1.9469&1.9323&&&1.926\\
&7&1.8957&1.9286&1.8512&&&&1.825\\
&8&1.8617&1.8656&1.7773&&&&1.748\\
&9&1.8320&1.8145&1.7180&&&&1.685\\
&10&1.8062&1.7721&1.6688&&&&1.634\\
&11&1.7150&1.6310&1.5080&&&&1.469\\
&20&1.6590&1.5479&1.4156&&&&1.373\\
&25&1.6201&1.4908&1.3531&&&&1.311\\
&50&1.5211&1.3430&&&&&\\
&100&1.44716&1.2253&&&&&\\
&500&1.3340&0.9812&&&&&\\
&1000&1.3007&0.9474&&&&&\\\tableline
\multirow{10}{2mm}{\rotatebox{90}{$n-6$\;}}&7&1.9894&2.1942&2.1798&&&&2.176\\
&8&1.9690&2.1310&2.1072&&&&2.102\\
&9&1.9464&2.0803&2.0499&&&&2.049\\
&10&1.9243&2.0386&2.0031&&&&1.997\\
&15&1.8352&1.9016&1.8506&&&&1.841\\
&20&1.7752&1.8213&&&&&\\
&50&1.6194&1.6246&&&&&\\
&100&1.5330&1.5123&&&&&\\
&500&1.3989&1.3103&&&&&\\
&1000&1.3593&1.2389&&&&&\\ \tableline
\multirow{10}{2mm}{\rotatebox{90}{$n-8$\;}}&10&1.9823&2.1834&2.1721&&&&\\
&15&1.9093&2.0524&2.02858&&&&\\
&20&1.8510&1.9769&&&&&\\
&25&1.8072&1.9253&&&&&\\
&50&1.6874&1.7932&&&&&\\
&100&1.5932&1.6895&&&&&\\
&500&1.4448&1.50266&&&&&\\
&1000&1.4008&1.4355&&&&&\\ \tableline
\multirow{2}{2mm}{\rotatebox{90}{\ \ $n-10$}}&\vrule height 4mm width 0mm depth 1mm 200&1.5606&1.7277&&&&&\\
&\vrule height 4mm width 0mm depth 2mm 500&1.4804&1.6363&&&&&\\ \tableline
SM&3&1.8928&1.8184&1.4677&1.2593&1.16867&&\\ 
\end{tabular}}
\end{ruledtabular}
\end{table}

Figs. \ref{tcdh1}, \ref{tcdh2} and \ref{tcdh3} show the plot of $v_c-{v_c}_{sq}$ respectively for $p=n-2$, $p=n-4$ and $p=n-6$ versus the segmentation step $k$. $v_c-{v_c}_{sq}$ fits quite reliably a power law of exponent $-k$:
\begin{equation}
v_c-{v_c}_{sq}=v_c-\sqrt{2}+1=a\left(1+b^{-k}\right)
\label{vcinf}
\end{equation}
where the parameters $a$ and $b$ depend of $n$ and $p$ and are listed in table \ref{tablevc}. The least square linear fits shown in dotted lines have an accuracy $R^2\leq10^{-5}$. However, no such power law is statisfied by $SM(3,1,k)$, the strict scale invariance seems necessary to get equation \ref{vcinf}.

\begin{center}
\includegraphics[width=8.5cm]{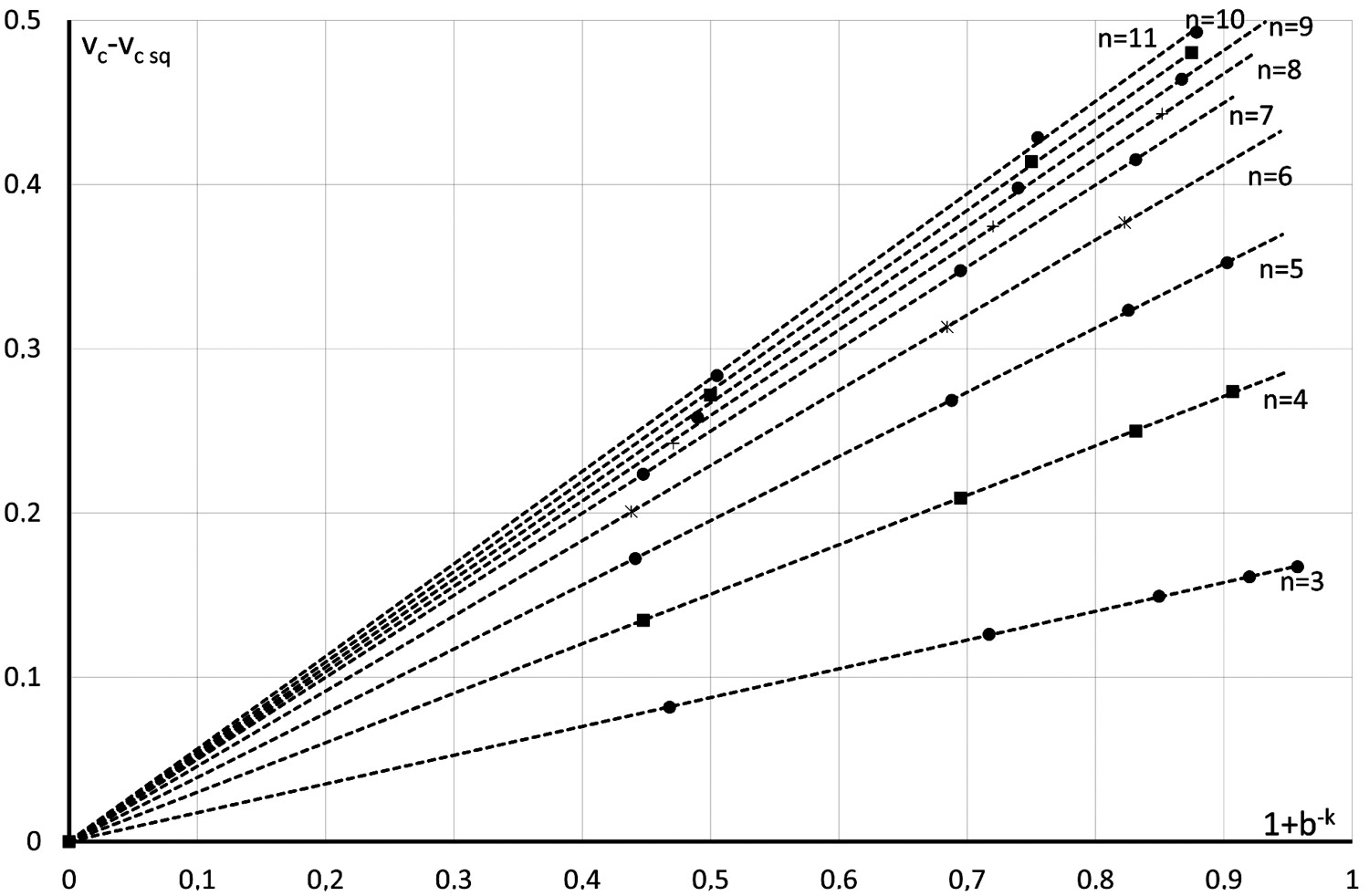}
\captionof{figure}{Plot of $v_c-{v_c}_{sq}$ versus $\left(1+b^{-k}\right)$ for $SC(n,n-2,k)$. The slopes $a$ and values of $b$ are given in table \ref{tablevc}.} 
\label{tcdh1}
\end{center}

\begin{center}
\includegraphics[width=8.5cm]{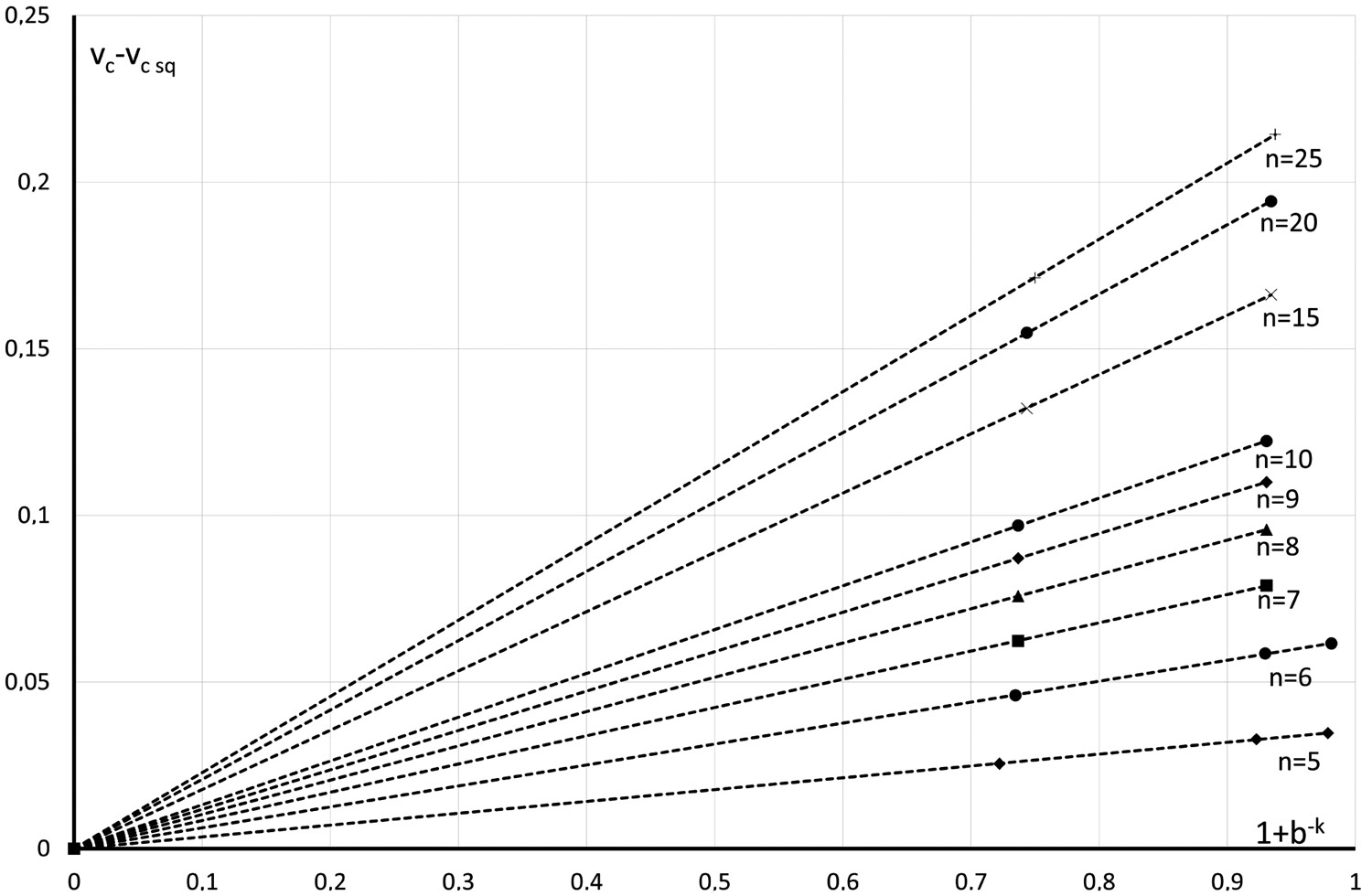}
\captionof{figure}{Plot of $v_c-{v_c}_{sq}$ versus $\left(1+b^{-k}\right)$ for $SC(n,n-4,k)$. The slopes $a$ and values of $b$ are given in table \ref{tablevc}.} 
\label{tcdh2}
\end{center}

\begin{center}
\includegraphics[width=8.5cm]{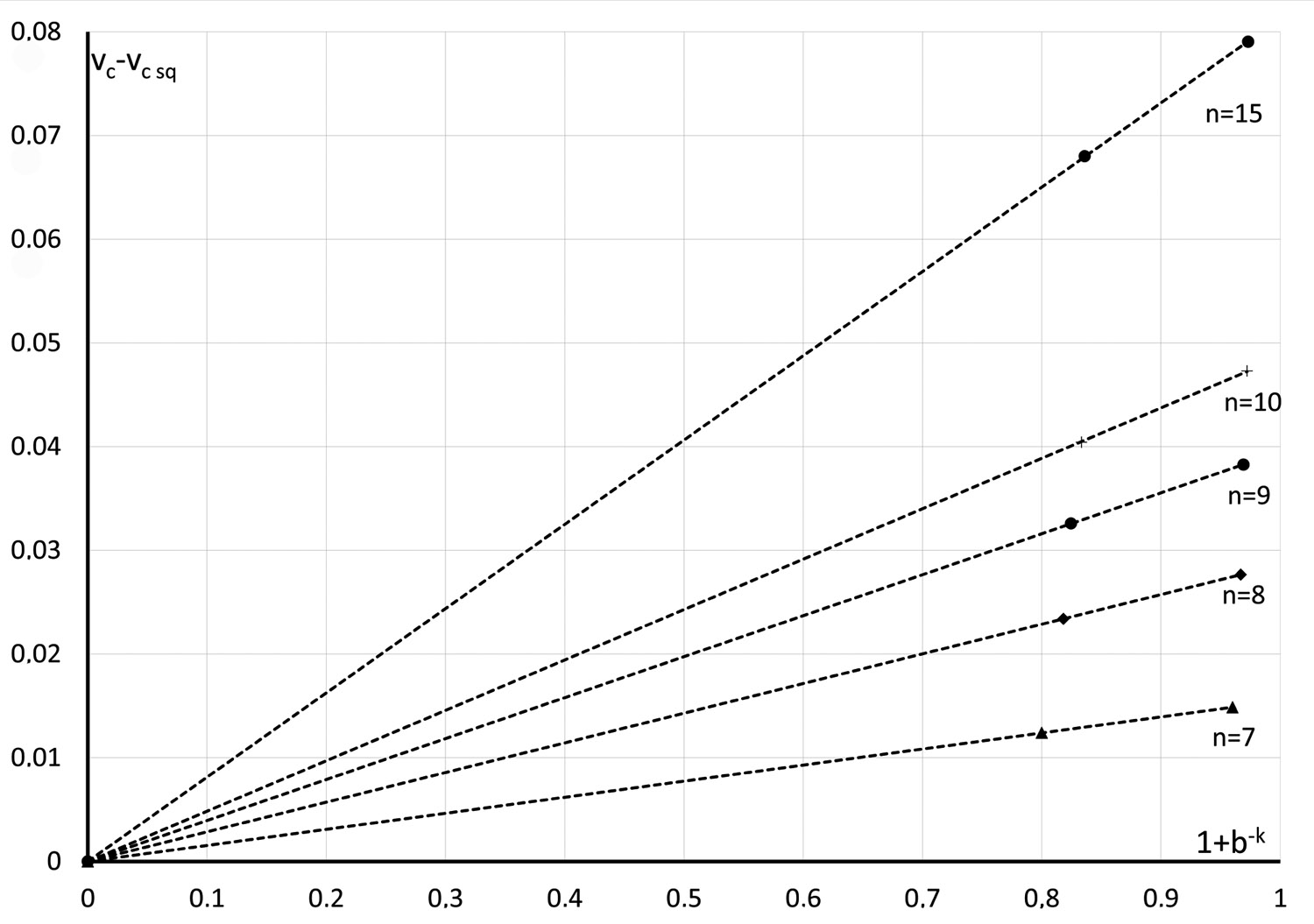}
\captionof{figure}{Plot of $v_c-{v_c}_{sq}$ versus $\left(1+b^{-k}\right)$ for $SC(n,n-6,k)$. The slopes $a$ and values of $b$ are given in table \ref{tablevc}.} 
\label{tcdh3}
\end{center}

\begin{table}[h]
\caption{Values of parameters $a$ and $b$ of the linear fits of Figs. \ref{tcdh1}, \ref{tcdh2} and \ref{tcdh3}.}
\label{tablevc}
\begin{ruledtabular}
\begin{tabular}{c|lcc}
&$SC$&$a$&$b$\\ \tableline
&SC(3,1)&0.1752&1.88\\
&SC(4,2)&0.3013&1.81\\
&SC(5,3)&0.3909&1.79\\
&SC(6,4)&0.4579&1.78\\
$p=n-2$&SC(5,3)&0.4996&1.81\\ 
&SC(8,6)&0.5204&1.89\\ 
&SC(9,7)&0.5365&1.96\\
&SC(10,8)&0.5501&2.00\\
&SC(11,9)&0.5633&2.02\\  \tableline
&SC(5,1)&0,0355&3.60\\
&SC(6,2)&0.0628&3.77\\
&SC(7,3)&0.0847&3.80\\
&SC(8,4)&0.1028&3.80\\ 
$p=n-4$&SC(9,5)&0.1182&3.80\\
&SC(10,6)&0.1314&3.80\\
&SC(15,11)&0.1778&3.90\\ 
&SC(20,16)&0.208&3.90\\
&SC(25,21)&0.2286&4.00\\ \tableline
&SC(7,1)&0.0155&5.00\\
&SC(8,2)&0.0286&5.50\\
$p=n-6$&SC(9,3)&0.0395&5.70\\ 
&SC(10,4)&0.0486&6.00\\
&SC(15,9)&0.0813&6.10\\ \end{tabular}
\end{ruledtabular}
\end{table}

Once $a$ and $b$ determined, the critical temperature for $k\rightarrow\infty$ $T_c(k=\infty)$ can be extrapolated from equation \eqref{vcinf}. The values are listed in the last column of table \ref{tab:tabletemp}. Fig. \ref{tcdhinf} shows the plot of ${\rm log}(T_c(\infty))$ versus $d_h$ for $p=n-2$, $p=n-4$ and $p=n-6$. The curves are approximately linear giving the following power laws:
\begin{equation}
\begin{array}{cl}
\hbox{for}\quad p=n-2,\qquad\qquad &T_c(\infty)\approx 2.48\,10^{-3}\ (30)^{d_h}\\
\hbox{for}\quad p=n-4,\qquad\qquad &T_c(\infty)\approx 0.184\ (3.32)^{d_h}\\
\hbox{for}\quad p=n-6,\qquad\qquad &T_c(\infty)\approx 0.344\ (2.49)^{d_h}\\
\end{array}
\end{equation}

\begin{center}
\includegraphics[width=8.5cm, height=6.5cm]{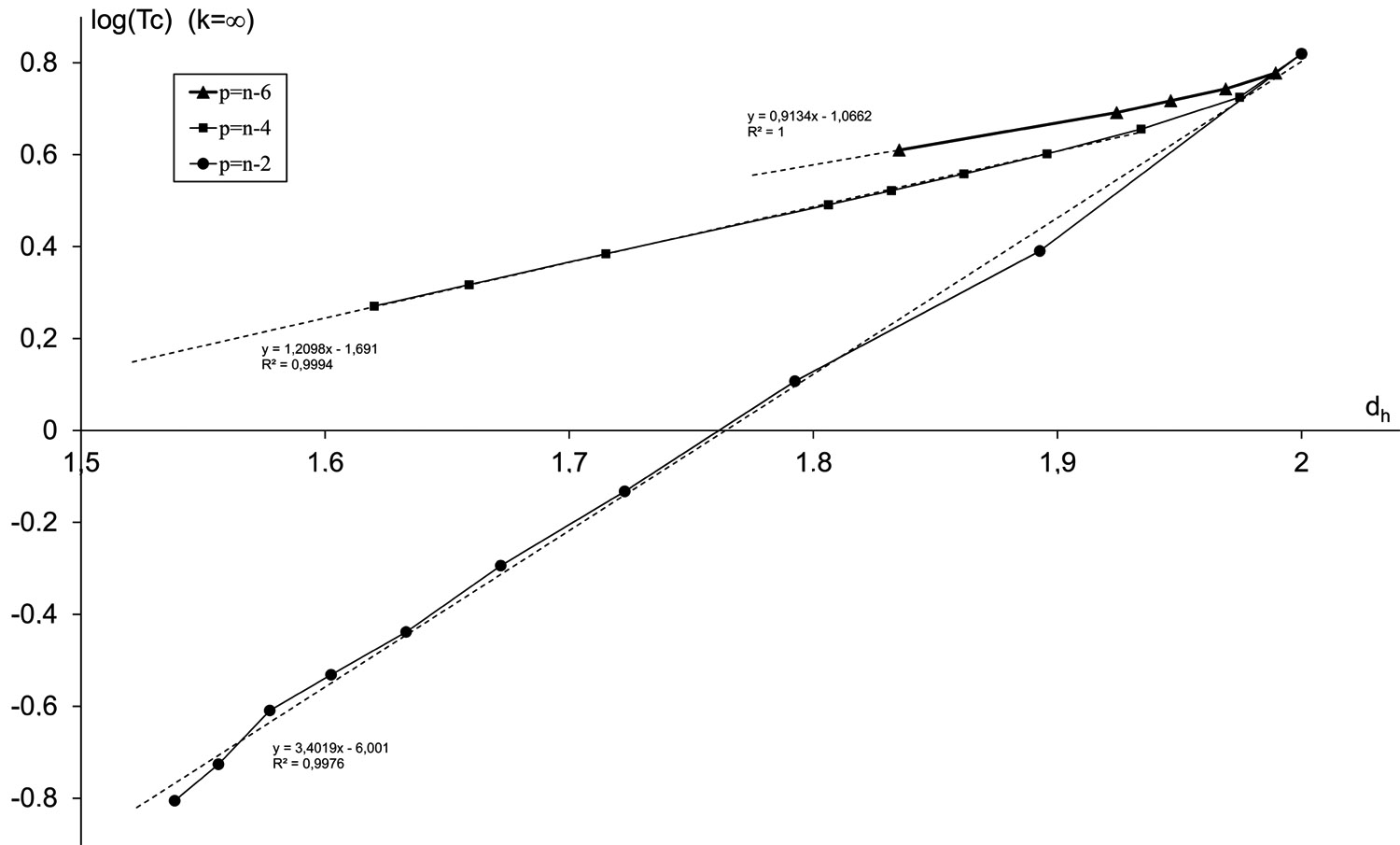}
\captionof{figure}{Plot of ${\rm log}(T_c(\infty))$ versus $d_h$. Dotted lines are linear fits.} 
\label{tcdhinf}
\end{center}

Fig. \ref{lnvc} shows the plot of the logarithm of $v_c$ versus $1/n$ for $SC(n,n-2,k)$. For $k=2$, the plot is approximately linear, at least for large values of $n$. A linear fit gives 
$$\log(v_c)\approx{-0.8132\over n}\quad(R^2=0.9998)$$

However, such a linear approximation does not hold for $k\neq 2$ or $p\neq n-2$.

\begin{center}
\includegraphics[width=8.5cm]{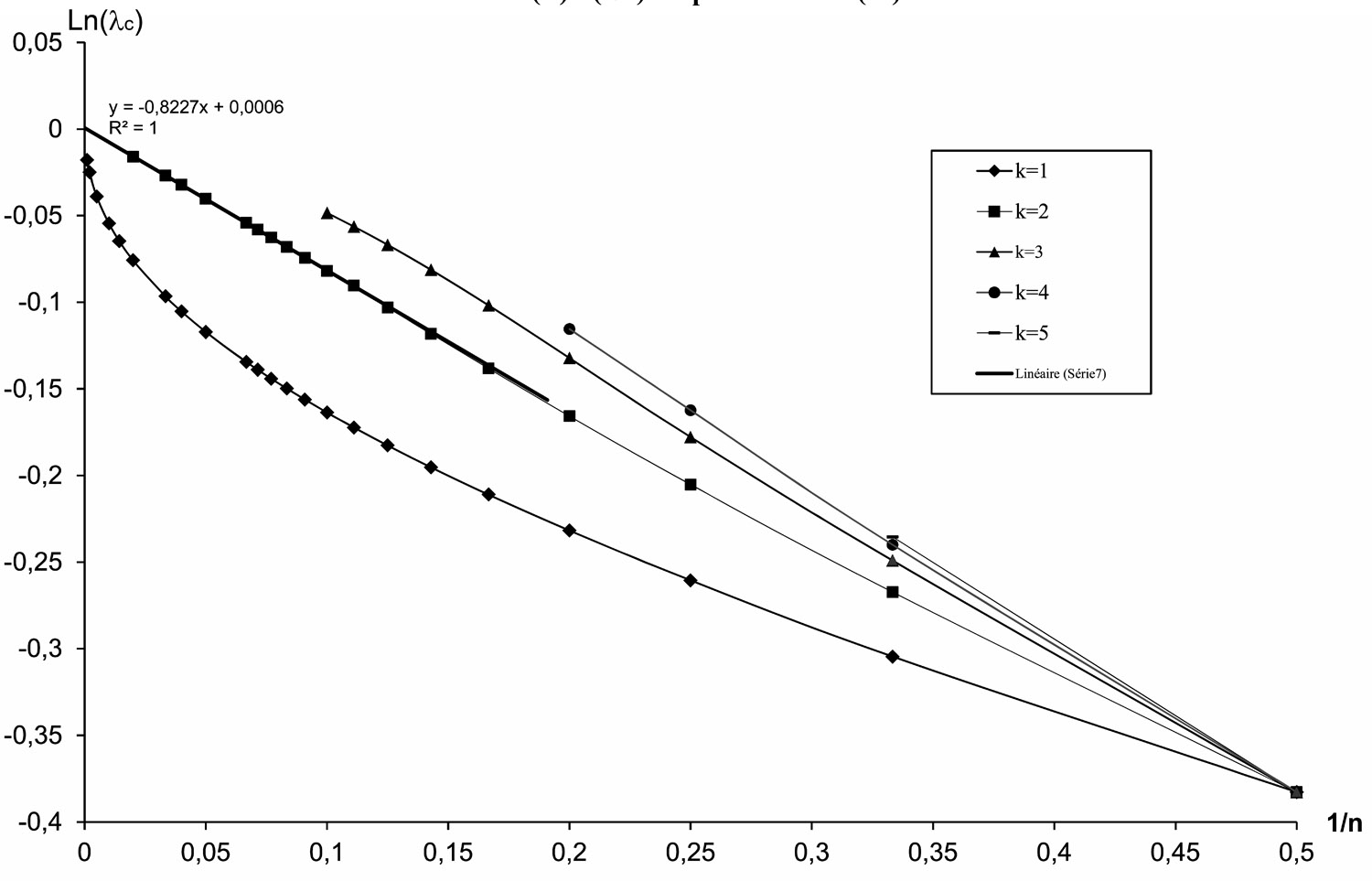}
\captionof{figure}{Plot of $\log(v_c)$ versus $1/n$, for $SC(n,n-2,k)$.} 
\label{lnvc}
\end{center}

In addtion to the critical temperature, the full spectrum of the pseudopartition function may be calculated. Several examples are plotted in the complex plan in appendix A. They may be compared to other fractal spectra obtained in hierarchical models\cite{derr}.

\section{Thermodynamics}

\subsection{Internal energy and specific heat on $SL(\infty,n)$}
The generating pattern of $SL(N,n)$ contains $2n-1$ sites and $2n$ bonds. The total number of sites is ${\cal N}=(2n-1)N^2$ and from (\ref{pseudoI2d}), the pseudopartition function of $SL(N,n)$ is
\begin{equation}
\begin{small}
\begin{aligned}
Z&=2^{\cal N}\cosh(\beta J)^{2nN^2}P_{N,n}(v)={2^{\cal N}P_{N,n}(v)\over (1-v^2)^{nN^2}}\\
&=2^{\cal N}\prod_{p,q=1}^{N}\left[R_1(n,v)-R_2(n,v)\left(\cos{2p\pi\over N}+\cos{2q\pi\over N}\right)\right]^{1\over 2}
\end{aligned}
\end{small}
\label{parti2ddilue}
\end{equation}
where
$$R_1(n,v)={(1+v^{2n})^2\over (1-v^2)^{2n}} \ \ \hbox{and}\ \  R_2(n,v)={2v^n(1-v^{2n})\over (1-v^2)^{2n}},$$
and calculations of the filled square Ising model are extended in a straight to the lacunary square lattice. The free energy $F$ is obtained taking the logarithme of $\eqref{parti2ddilue}$, dividing by $\cal N$ and, at the limit $N\rightarrow\infty$, replacing the double sum by a double integral:
\begin{equation}
\begin{aligned}
\beta F=-\log 2-&{1\over 8(2n-1)\pi^2}\iint_0^{2\pi}\log\left[R_1(n,v)\right.\\ 
&-\left. R_2(n,v)(\cos{x}+\cos{y})\right]dxdy
\end{aligned}
\end{equation}

Differentiating a first time, we obtain the internal energy $U={\partial \beta F\over\partial \beta}$. Let us note $R'_1$ and $R'_2$ the derivatives of $R_1$ and $R_2$ with respect to $v$:
\begin{widetext}
\begin{equation*}
\begin{small}
\begin{aligned}
U=&-{1\over 8(2n-1)\pi^2}{dv\over d\beta}\iint_0^{2\pi}{R'_1-R'_2\left(\cos{x}+\cos{y}\right)\over R_1-R_2\left(\cos{x}+\cos{y}\right)}dxdy=-{(1-v^2)\over 128}\Biggl[{R'_2\over R_2}+{2\over \pi}\left({R'_1\over R_1}-{R'_2\over R_2}\right)K\left({2R_2\over R_1}\right)\Biggl]\\
=&{n\over (2n-1)}\Biggl[{\left(v^{2 n+2}+3 v^{2 n}-3 v^2-1\right)\over 2 v \left(1-v^{2 n}\right)}+\frac{\left(1-v^2\right) \left(v^{4 n}-6 v^{2 n}+1\right)}{\pi v(1-v^{4n})}K\left(\frac{4 v^n \left(v^{2 n}-1\right)}{\left(v^{2 n}+1\right)^2}\right)\Biggl]
\end{aligned}
\end{small}
\end{equation*}
where $K$ is the elliptic integral of the first kind. $U$ is plotted in Fig. $\ref{plotUi2D}$ for $n=1$ to $5$ and $n=10$.

Differentiating a second time, we obtain the specific heat $C=-\beta^2{\partial^2 \beta F\over\partial \beta^2}$.
\begin{equation}
\begin{small}
\scriptsize
\begin{aligned}
C=\frac{n\beta^2}{2(2n-1)\pi}&\Biggl[\frac{\left(2 \left(v^2+1\right) \left(v^{6 n}-7 v^{4 n}+7 v^{2 n}-1\right)+n \left(v^2-1\right) \left(v^{6 n}+11 v^{4 n}+11 v^{2 n}+1\right)\right)}{ v^2 \left(v^{2 n}-1\right)^2 \left(v^{2 n}+1\right)} K\left(\frac{4 v^n \left(1-v^{2
  n}\right)}{\left(v^{2 n}+1\right)^2}\right)\\
  & -\frac{n\left(v^2-1\right)\left(v^{2 n}-2 v^n-1\right)^2}{ v^2 \left(v^{2 n}-1\right)^2} E\left(\frac{4 v^n\left(1-v^{2 n}\right)}{\left(v^{2 n}+1\right)^2}\right)-\frac{\pi\left(4 (n-1) v^{2 n+2}+v^{4 n+2}-3 v^{4 n}-4 (n-1) v^{2 n}+3 v^2-1\right)}{v^2 \left(v^{2 n}-1\right)^2}\Biggl]
\end{aligned}
\end{small}
\end{equation}
\end{widetext}
where $E$ is the elliptic integral of the second kind. $C$ is plotted on Fig. $\ref{plotCi2D}$ for $n=1$ to $5$ and $n=10$.

\begin{center}
\includegraphics[width=\hsize]{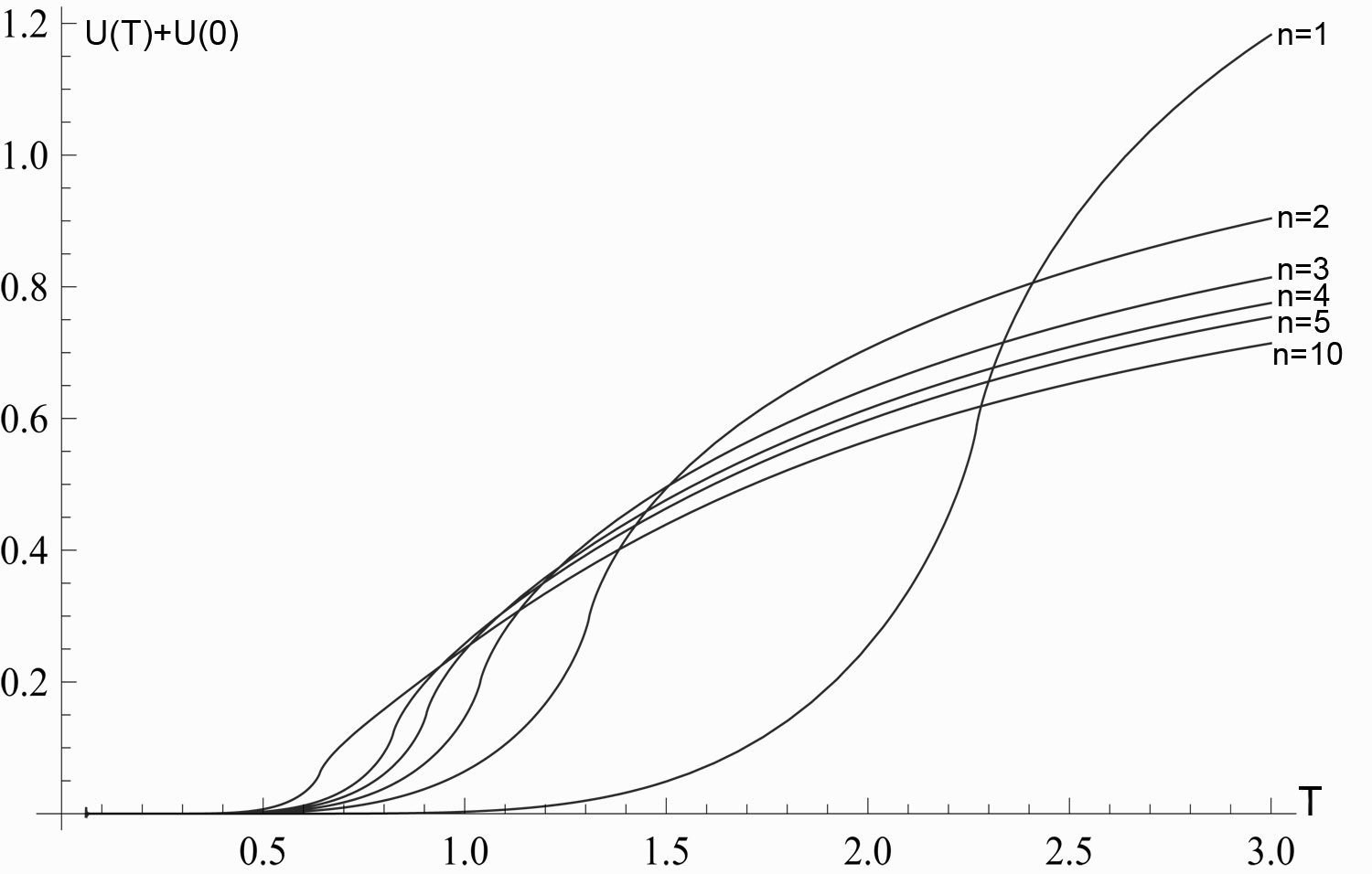}
\captionof{figure}{The internal energy $U$ of $SL(\infty,n)$ versus temperature for $n=1$ to $5$ and $10$.} 
\label{plotUi2D}
\end{center}

\begin{center}
\includegraphics[width=\hsize]{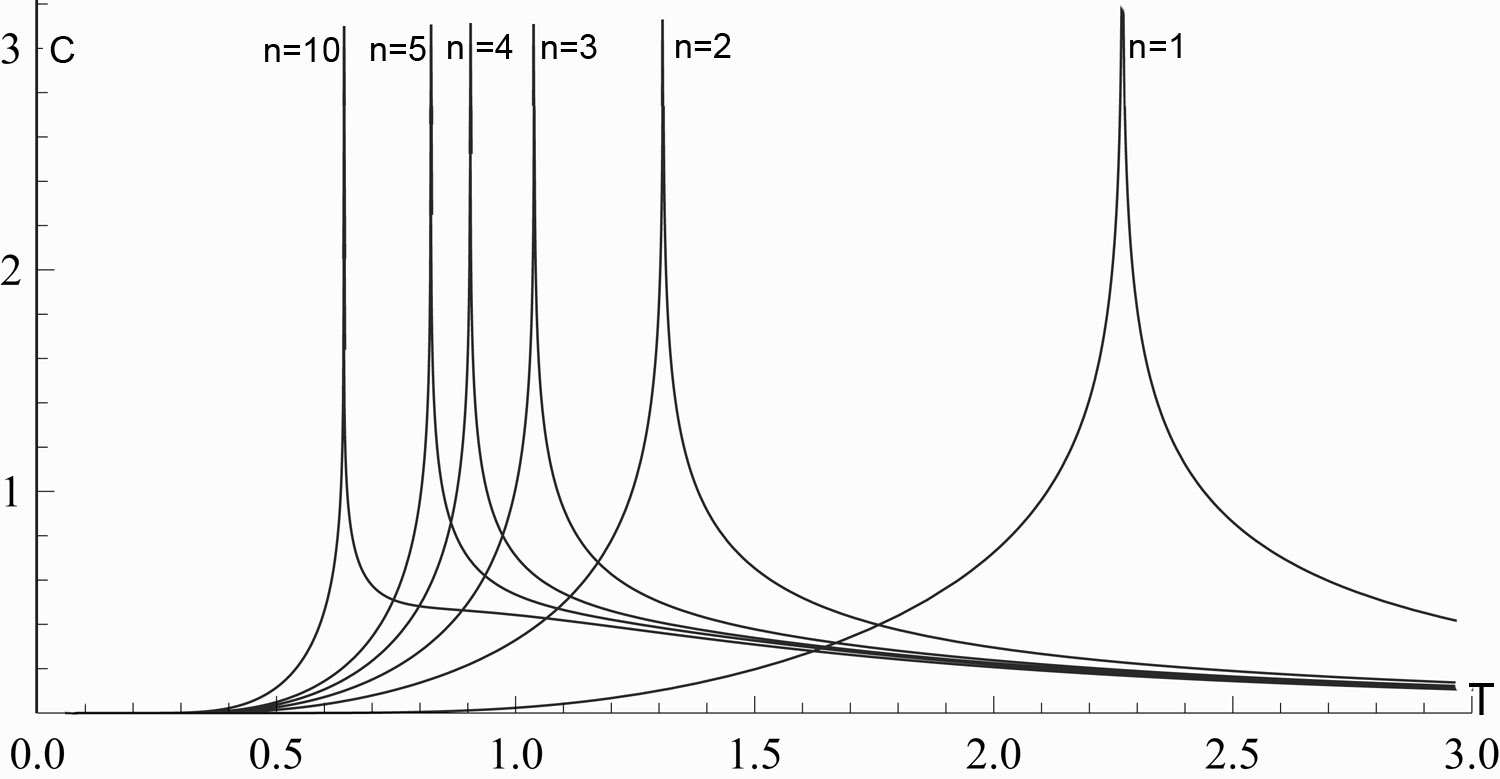}
\captionof{figure}{The specific heat $C$ of $SL(\infty,n)$ versus temperature for $n=1$ to $5$ and $10$.} 
\label{plotCi2D}
\end{center}

\subsection{Internal energy and specific heat on Sierpi\'nski carpets}

The generating pattern of $SC(3,1,2)$ has $8$ sites and $14$ bonds, then the pseudopartition function of $SC(3,1,1)$ is
$$Z\ =\ 2^{8N^2}\ {P_{N,SC(3,1,1)}(v)\over (1-v^2)^{7N^2}}$$

From \eqref{pseudosier} and taking the limit $N\rightarrow\infty$, the free energy per site is
\begin{equation}
\begin{small}
\begin{aligned}
\beta F=-\log 2-{1\over 64\pi^2}\iint_0^{2\pi}\log\left[R_1(v)+R_2(v)(\cos x\right.\qquad\\
+\cos y)\left.+R_3(v)\cos x\cos y+R_4(v)(\cos^2x+\cos^2y)\right]dx dy
\end{aligned}
\end{small}
\label{energysc}
\end{equation}
where $\qquad\qquad\displaystyle{R_i(v)_{(1\leq i\leq 4)}={Q_i(v)\over (1-v^2)^{14}}}$

\begin{center}
\includegraphics[width=\hsize]{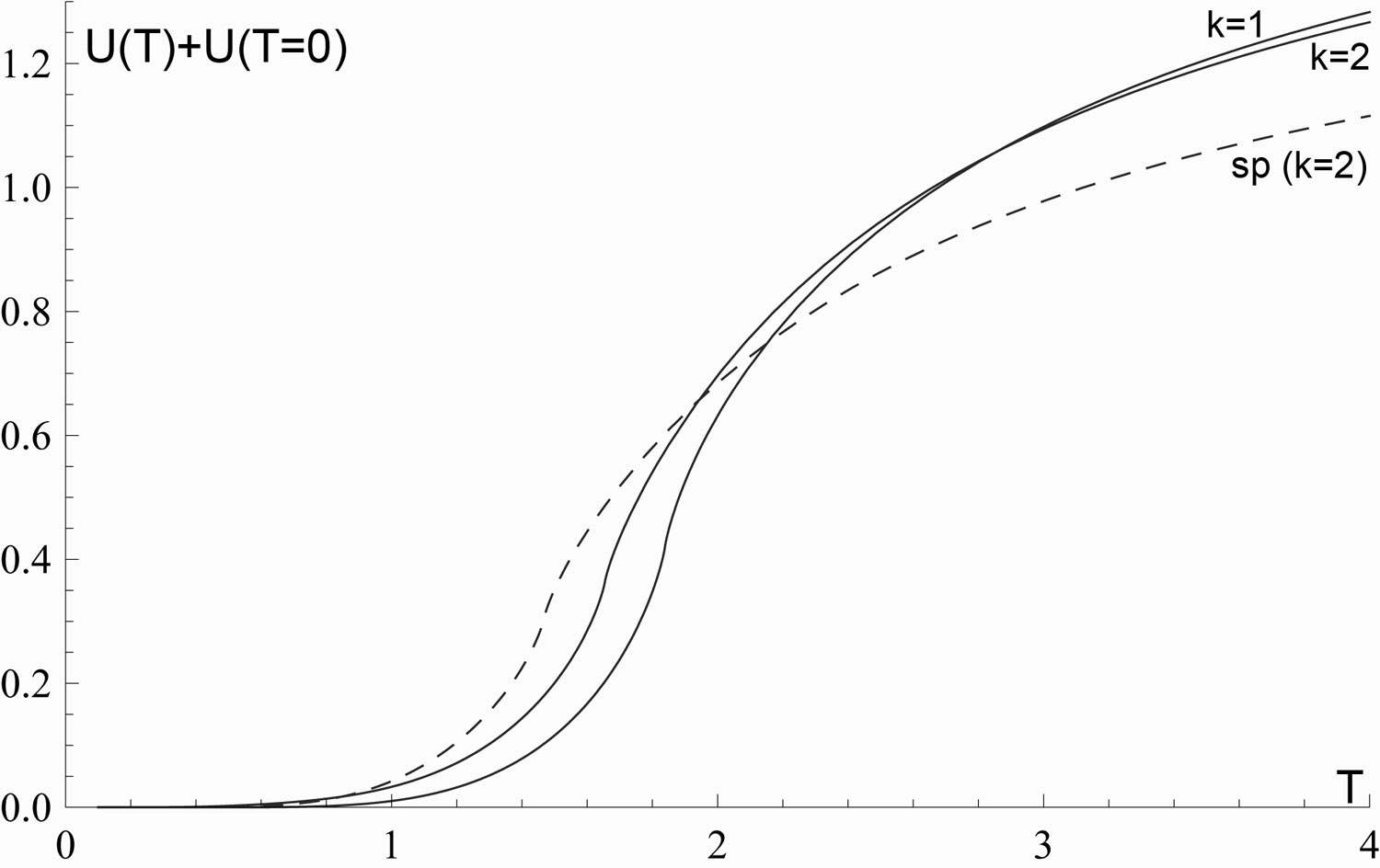}
\captionof{figure}{The internal energy of $SC(3,1,1)$, $SC(3,1,2)$ and $SM(3,2)$ (in dashed line) versus temperature.} 
\label{plotsierU}
\end{center}

Calculations have been performed also for the second segmentation steps $SC(3,1,2)$ and $SM(3,2)$. Their pseudopartition polynomials have nine terms and there degrees  are respectively $84$ and $64$. They are explicitely given in appendices B and C. Their internal energy and specific heat of $SC(3,1,1)$, $SC(3,1,2)$ and $SM(3,2)$ are plotted in Figs. $\ref{plotsierU}$ and $\ref{plotsierC}$.

\begin{center}
\includegraphics[width=\hsize]{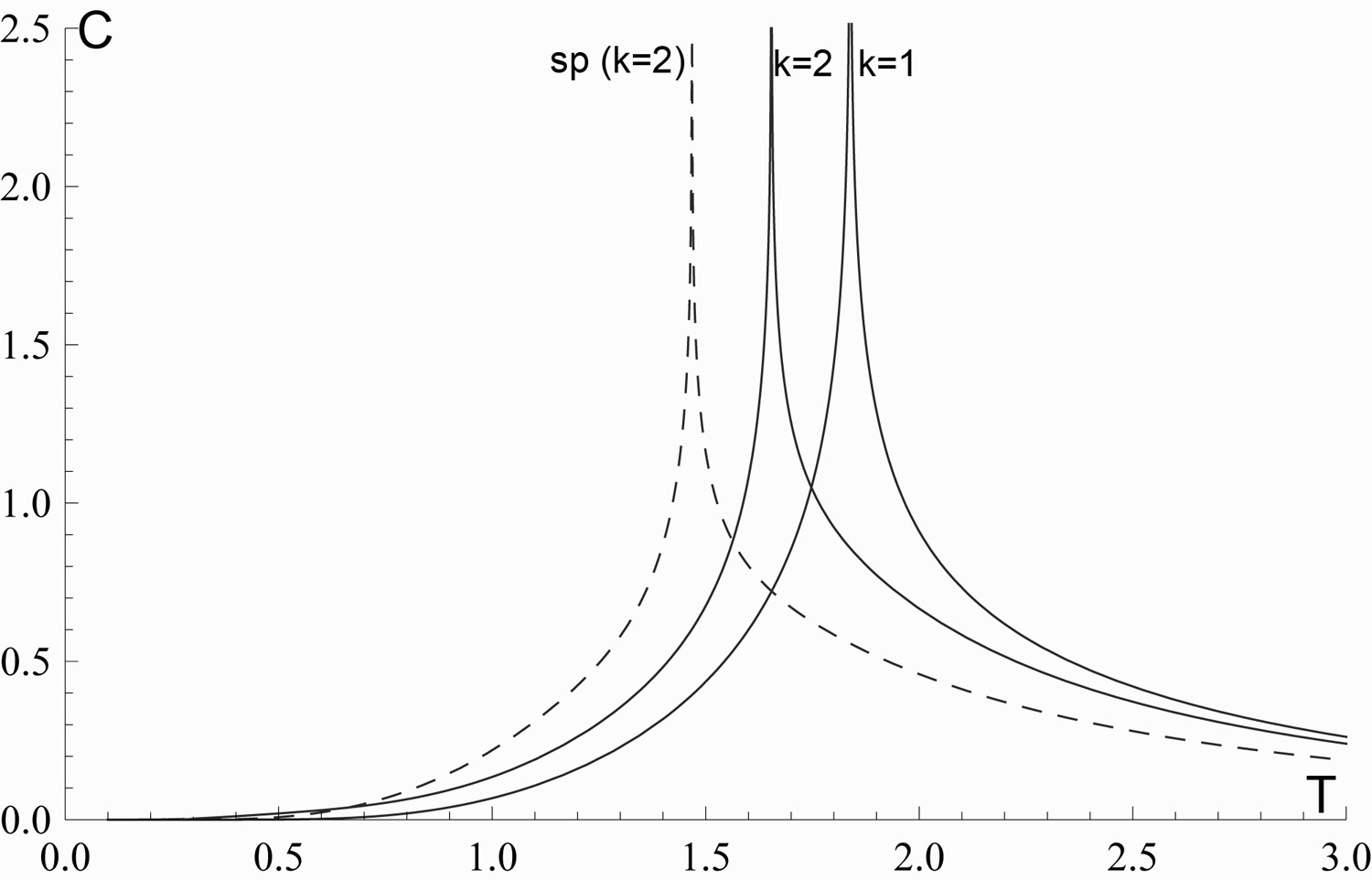}
\captionof{figure}{The specific heat of  $SC(3,1,1)$, $SC(3,1,2)$ and $SM(3,2)$ (in dashed line) versus temperature.} 
\label{plotsierC}
\end{center}

\subsection{Magnetization of the Ising model on $SL(\infty,n)$}
The magnetization $m$ of the square lattice without an external field has been explicitely calculated in \cite{yang}. The method has been subsequently refined \cite{mont, rum} and adapted to the loops counting method we are using in this paper\cite{pottsw}. The basic principles is to evaluate the asymptotic limit of average correlations between two spins when their distance become infinite. The calculation is generally performed within a row of spins:
$$m=\lim_{k\rightarrow\infty}<s_1,s_{k+1}>$$

In terms of the matrices $\cal M$ involved in the calculations of the partition function, correlations within one row of sites are calculated from matrices elements of $(I-v{\cal M})^{-1}$. 

For $SQ(N,n)$, using the translation invariance \cite{newmon} and the further reduction already used for the calculation of the partition polynomial, the matrix involved is the inverse of the the matrix of equation $\eqref{matred}$, replacing $2\pi p\over N$ and $2\pi q\over N$ respectively by $x$ and $y$, $M(N,p,q)\rightarrow M(x,y)$ (expressions are taken for $N\rightarrow\infty$):
\begin{equation}
\setlength{\arraycolsep}{0.3pt}
\begin{small}
\begin{array}{l}
\Delta(v,x,y)\left[I-v^nM(x,y)\right]^{-1}=\\ \noalign{\vglue4pt}
\begin{pmatrix}
\gamma_1(v,x,y)&\gamma_{2-}(v,-x,y)&\gamma_{2+}(v,-x,-y)&\gamma_3(v,-x,-y)\\
\gamma_{2+}(v,-y,x)&\gamma_1(v,y,x)&\gamma_3(v,-y,x)&\gamma_{2-}(v,-y,-x)\\
\gamma_{2-}(v,y,x)&\gamma_3(v,y,-x)&\gamma_1(v,-y,x)&\gamma_{2+}(v,y,-x)\\
\gamma_3(v,x,y)&\gamma_{2+}(v,x,y)&\gamma_{2-}(v,x,-y)&\gamma_1(v,-x,y)
\end{pmatrix}
\end{array}
\end{small}
\end{equation}
where
\begin{equation}
\setlength{\arraycolsep}{0.3pt}
\begin{array}{lcl}
\Delta(v,x,y)&=&\left|I-v^nM(x,y)\right|\\ 
&=&(1+v^{2n})^2-2v^n(1-v^{2n})(\cos x+\cos y)\\ \noalign{\vglue3pt}
\gamma_1(v,x,y)&=&1+v^{2n}-v^n(1-v^{2n})e^{ix}-2v^n\cos y\\  \noalign{\vglue3pt}
\gamma_{2\pm}(v,x,y)&=&e^{\pm{i\pi\over 4}}v^n(e^{ix}-v^n-v^ne^{ix+iy}-v^{2n}e^{iy})\\  \noalign{\vglue3pt}
\gamma_3(v,x,y)&=&2v^{2n}e^{ix}\sin y
\end{array}
\end{equation}

From these expressions, similar calculations as for $n=1$ \cite{pottsw, mont}, with the substitution $v\rightarrow v^n$ leads to the magnetization:

$$m=\left[1-{\left(1-v^{2n}\right)^4\over 16 v^{4n}}\right]^{1\over 8}$$
which is plotted on Fig. \ref{magdilsquare} for $n=1$ to $5$ and $n=10$.

\begin{center}
\includegraphics[width=\hsize]{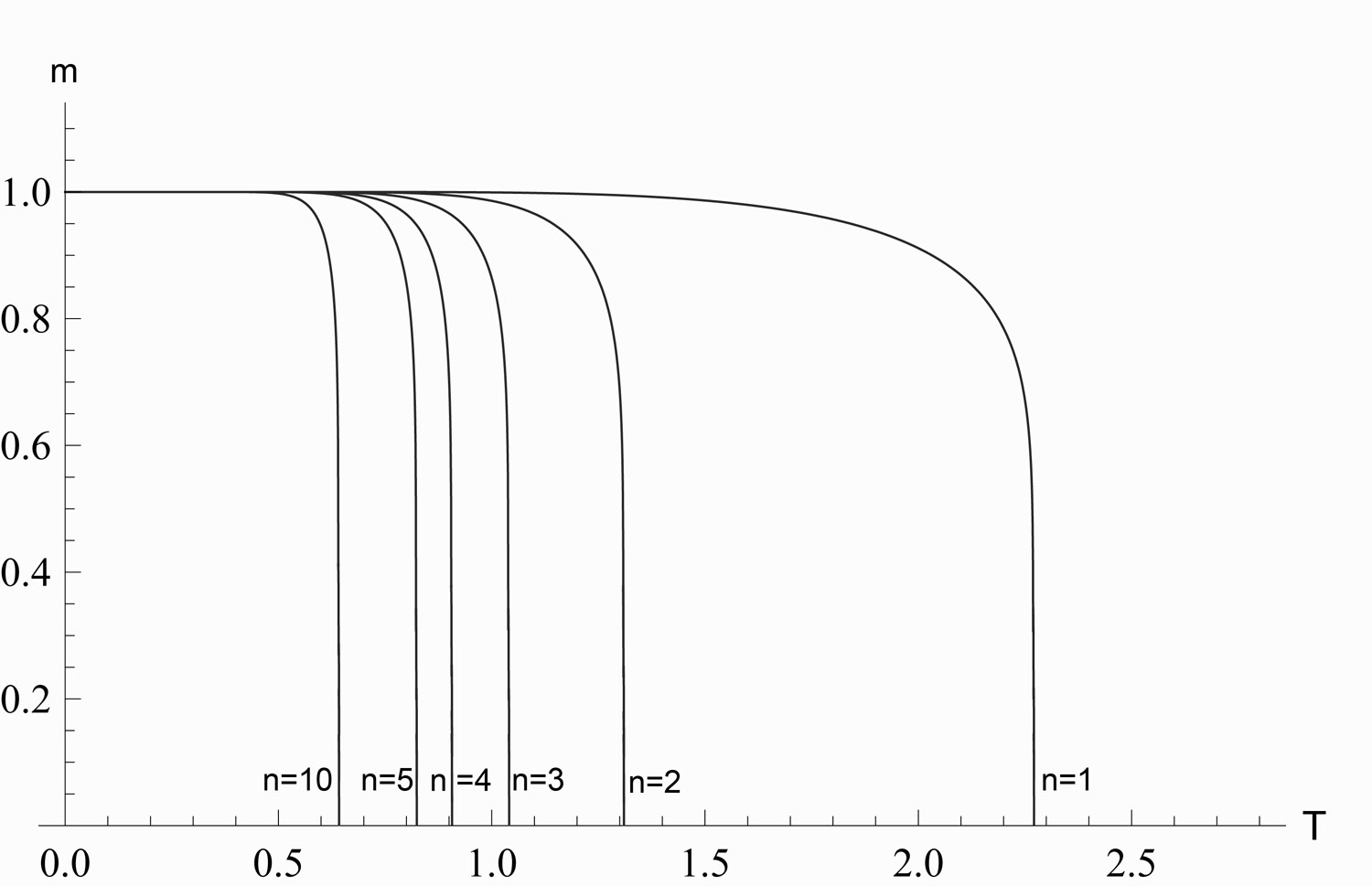}
\captionof{figure}{The magnetization of the lacunary square lattice $SQ(\infty,n)$, for $n=1$ to $5$ and $n=10$.} 
\label{magdilsquare}
\end{center}

When taking the first term of the expansion of $m$ in the vicinity of the critical parameter $v_c=(\sqrt{2}-1)^{1\over n}$, the magnetization writes
$$m^8={4n\sqrt{2}(\sqrt{2}+1)^{1\over n}(v-v_c)}+O(v-v_c)^2$$

However, the inversion of the matrix for Sierpi\'nsky carpets doesn't lead to straightforward calculations and we have not yet reached a closed formula for the magnetization.

\section{Discussion and conclusion}

\begin{table*}[t]
\caption{The critical temperatures of the Ising model on $SC(3,1,k)$ for finite values of $k$. Values which are between two columns correspond to evaluations obtained from crossing of two successive Binder cumulants.}
\label{tabres}
\begin{ruledtabular}
\begin{tabular}{cccccccccccc}
\setlength{\tabcolsep}{0pt}
&$k=1$&$k=2$&$k=3$&$k=4$&&$k=5$&&$k=6$&&$k=7$&$k=\infty$\\ \tableline
Monceau \& Perreau\cite{mon01}&&&&&$1.486$&&$1.482$&&$1.480$&&1.4795(5)\\
Carmona \& al.\cite{carm}&&&$1.724$&1.590&&1.538&&1.511&&1.497&1.481\\
Pruessner \& al.\cite{pruslois}&&&&1.5266(11)&&1.5081(12)&&1.4992(11)&&\\ 
This work&1.8384&1.6538&1.576&1.5256&&1.5045&&&&&1.478\\
\end{tabular}
\end{ruledtabular}
\end{table*}

At this step, we are facing the problem of taking the thermodynamic limit in fractal systems. Sierpi\'nsky carpets are obtained by the iteration of a segmentation rule step by step (which is noted $k$ in this paper). But at each finitie step, reproducing the generating pattern periodically $N^2$ times restore an artificial translation invariance. In a strictly meaning, thermodymical limit should be taken setting $k\rightarrow\infty$. That is done in this paper in section IV for the calculation of the critical temperature. But for a finite segmentation step $k$, calculations of thermodynamical functions (section V) have been done by setting $N\rightarrow\infty$. In this way, basic symmetries are the same as the square lattice and this is not surprizing to obtain the same critical exponents which obviously results from equations \eqref{energysc}, \eqref{partitionsca} and \eqref{partitionscb}. 

Then the comparison with simulations can only be done on critical temperatures. Table \ref{tabres} summarizes the critical temperatures of $SC(3,1)$ for finite values of $k$ and their extrapolations to $k\rightarrow\infty$ from litterature (the few references which give the detailed values for each investigated finite $k$) and from this work. Values obtained by Pruessner \& al.\cite{pruslois} are the most accurate with a difference less than $0.27\%$. This is coherent since it is the single reference which uses multiple translationally invariant reproductions of a generating pattern. However these authors do not give explicitely an extrapolation for $k\rightarrow\infty$ (in table \ref{tableref} the value for $k=6$ has been taken as an approximative value of $T_c(k=\infty)$ which is clearly overestimated). For $T_c(k=\infty)$, references obtaining results close to $1.48$\cite{mon98, carm, mon01, hsiao03} are more accurate that those obtaining results close from $1.50$\cite{bab1, bab3}.

To compare critical exponents with simulation as we are doing for critical temperatures, scale invariance should be implemented in order to get a fully solvable model of the Ising ferromagnetism on fractal lattices at the limit $k\rightarrow\infty$ which still remains an opened problem.

\appendix

\section{Spectra of partition functions of the Ising model on various Sierpi\'nski carpets}

\begin{center}
\includegraphics[width=\hsize]{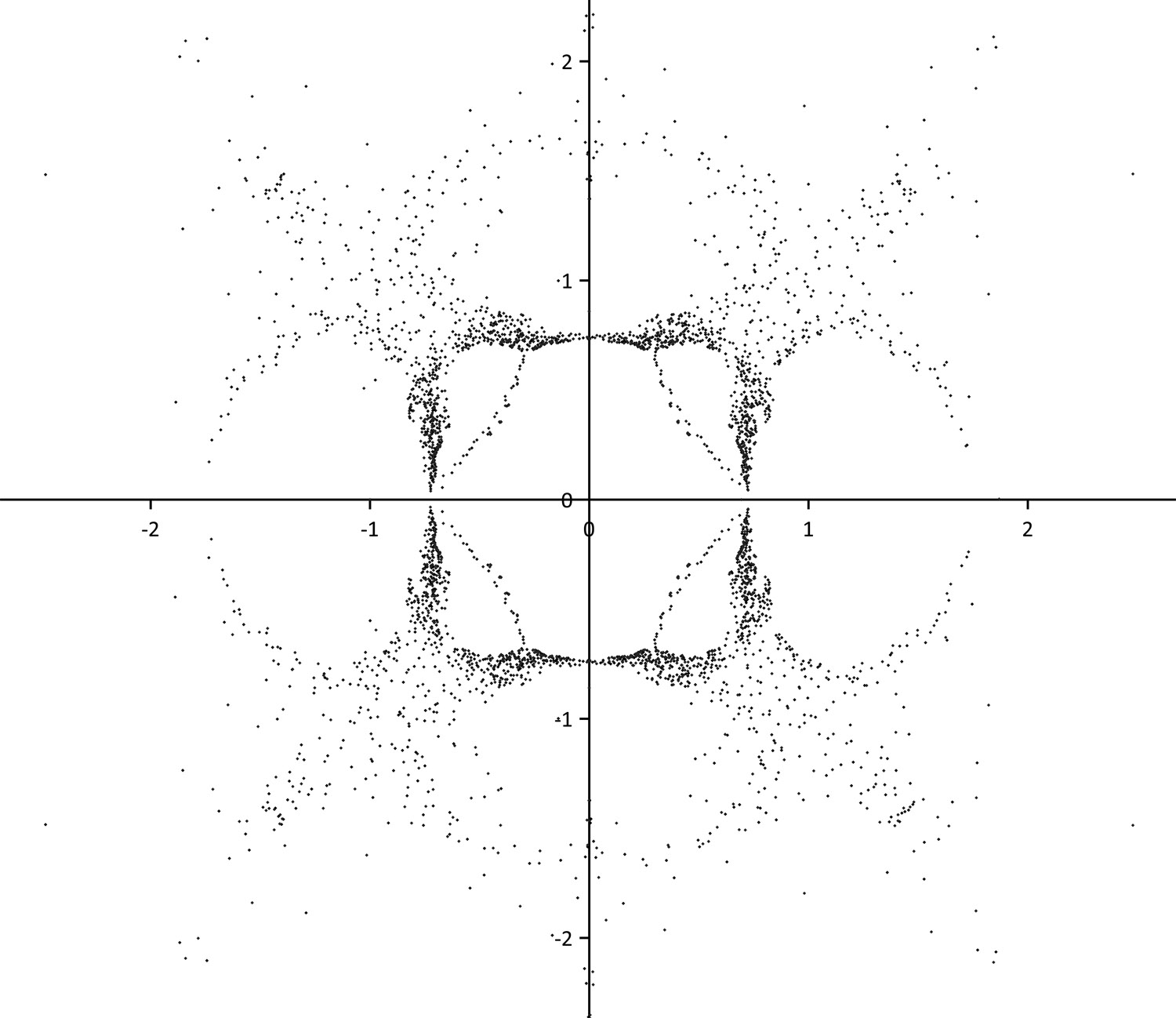}
\captionof{figure}{Spectrum of the Ising model on $SM(3,4)$.} 
\label{sspecial}
\end{center}

\begin{center}
\includegraphics[width=\hsize]{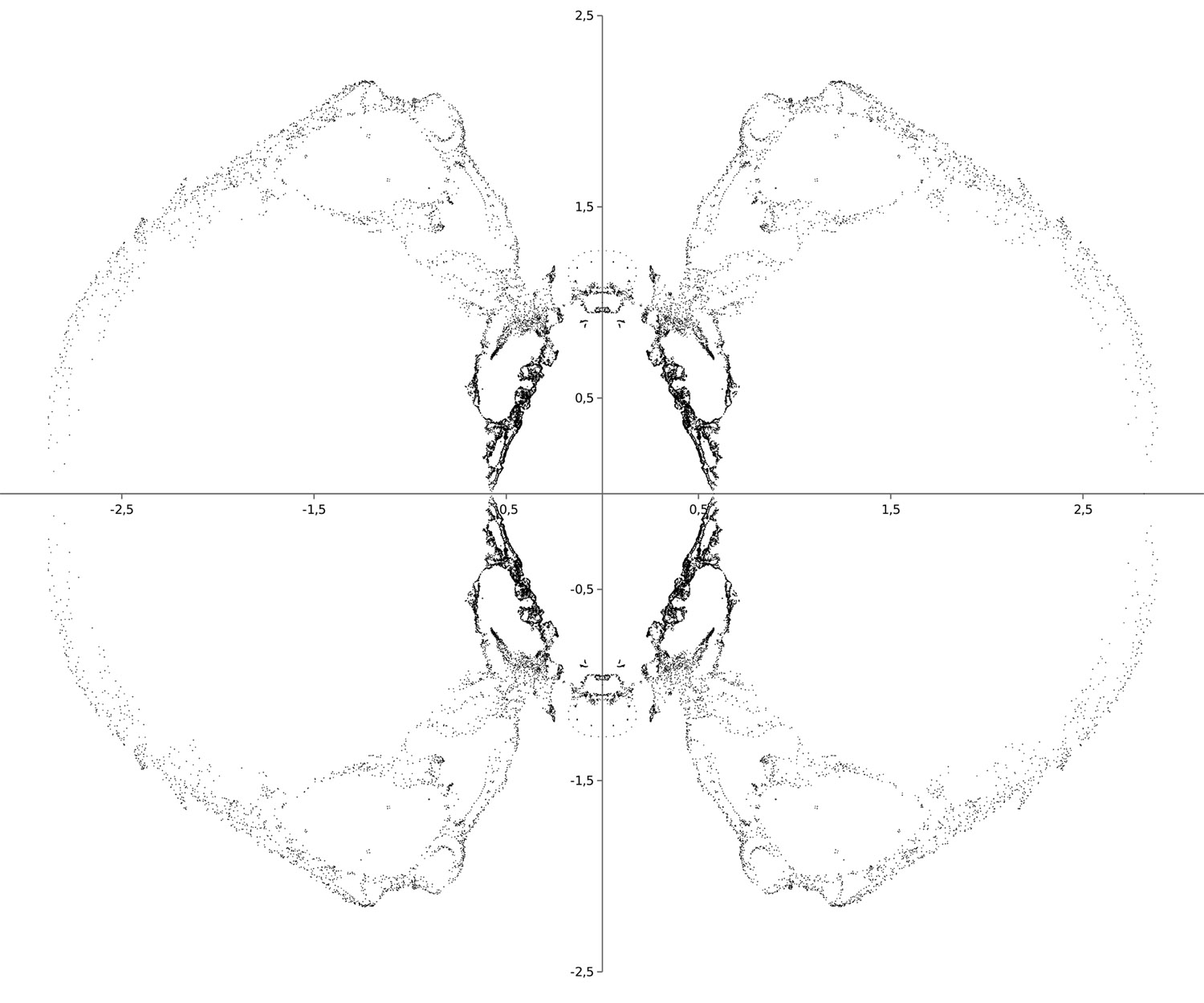}
\captionof{figure}{Spectrum of the Ising model on $SC(3,1,5)$.} 
\label{specn3p1}
\end{center}

\begin{center}
\includegraphics[width=\hsize]{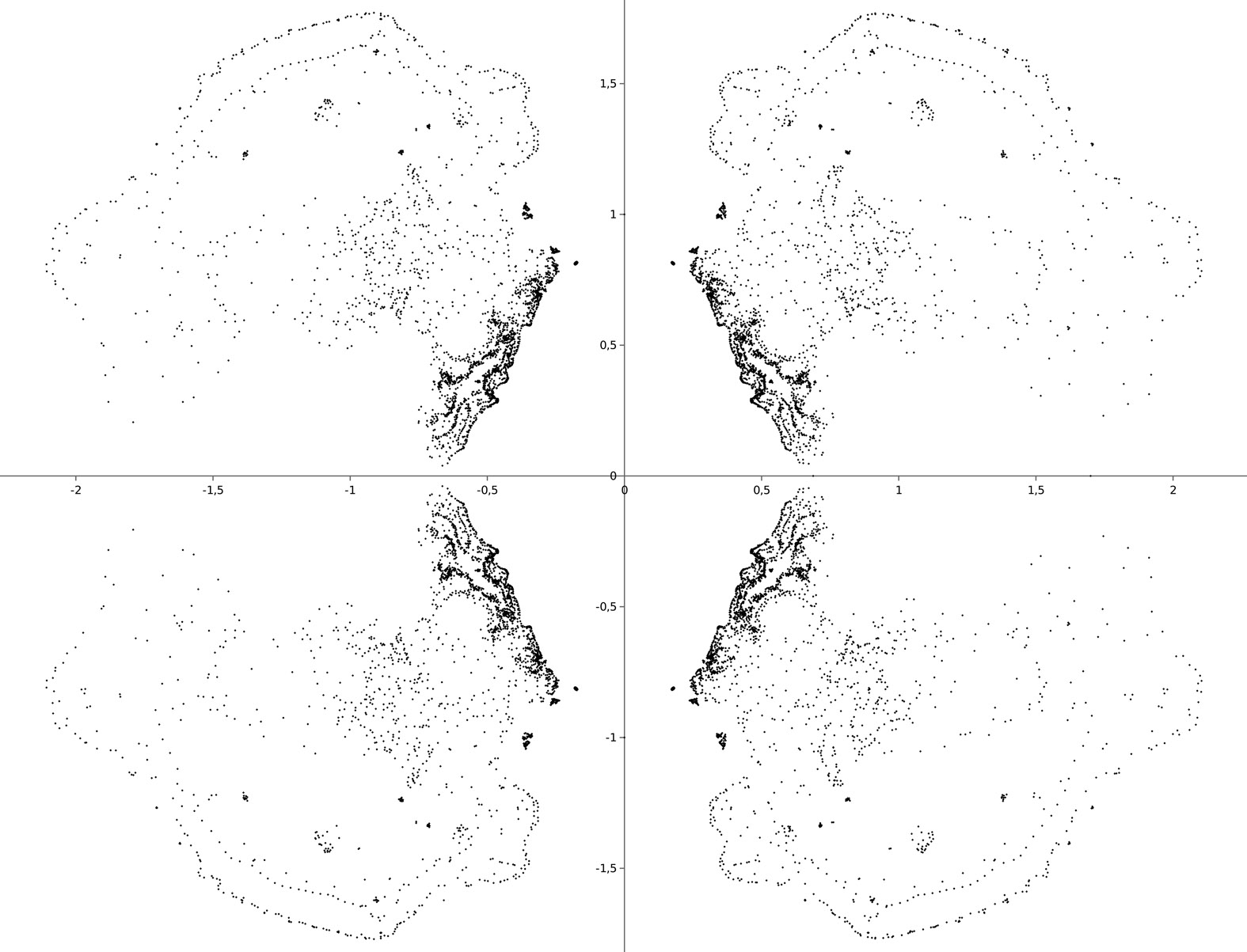}
\captionof{figure}{Spectrum of the Ising model on $SC(4,2,4)$.} 
\label{specn4p2}
\end{center}
 
\begin{center}
\includegraphics[width=\hsize]{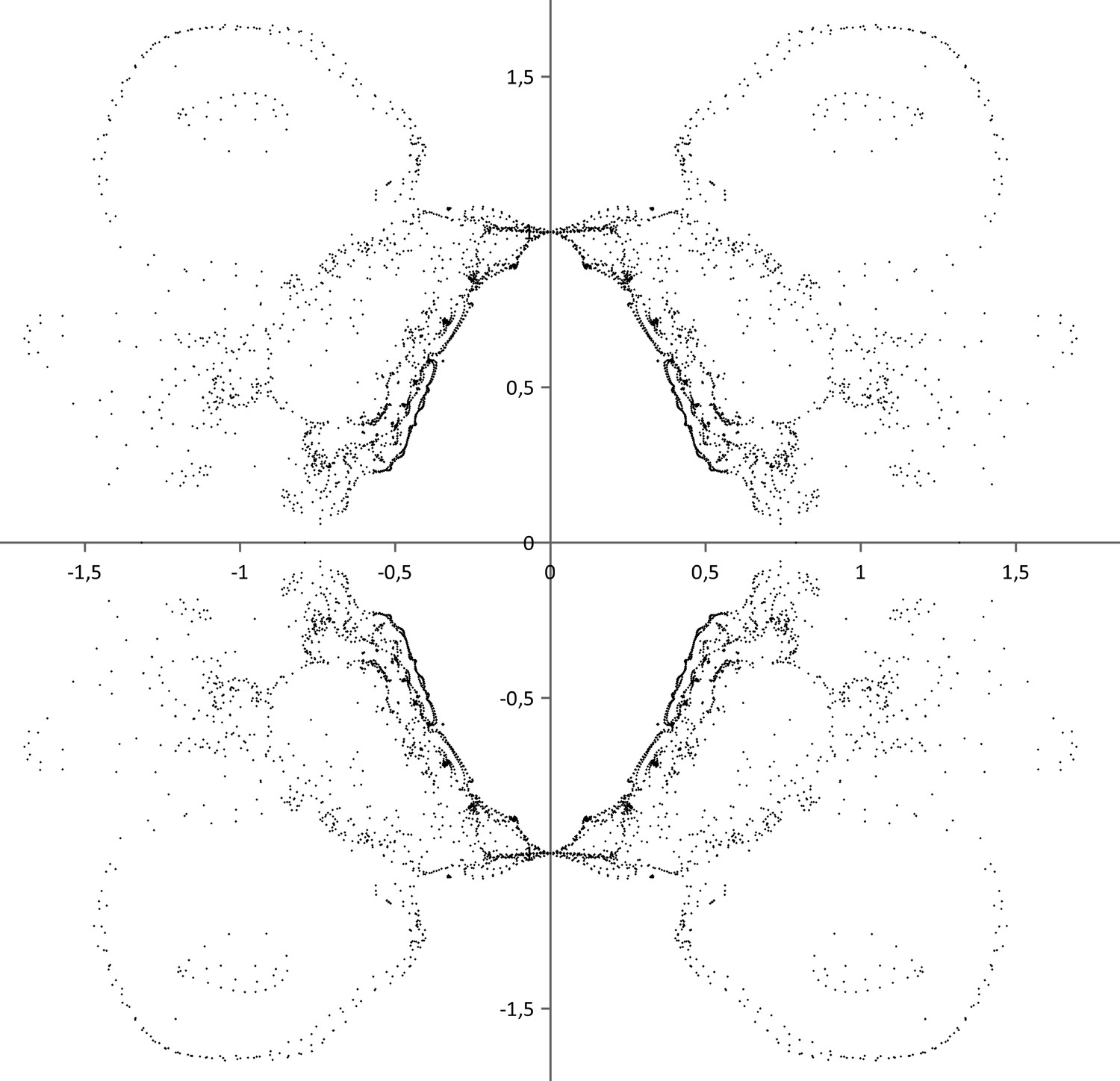}
\captionof{figure}{Spectrum of the Ising model on $SC(6,4,3)$.} 
\label{specn5p3}
\end{center}

\begin{center}
\includegraphics[width=\hsize]{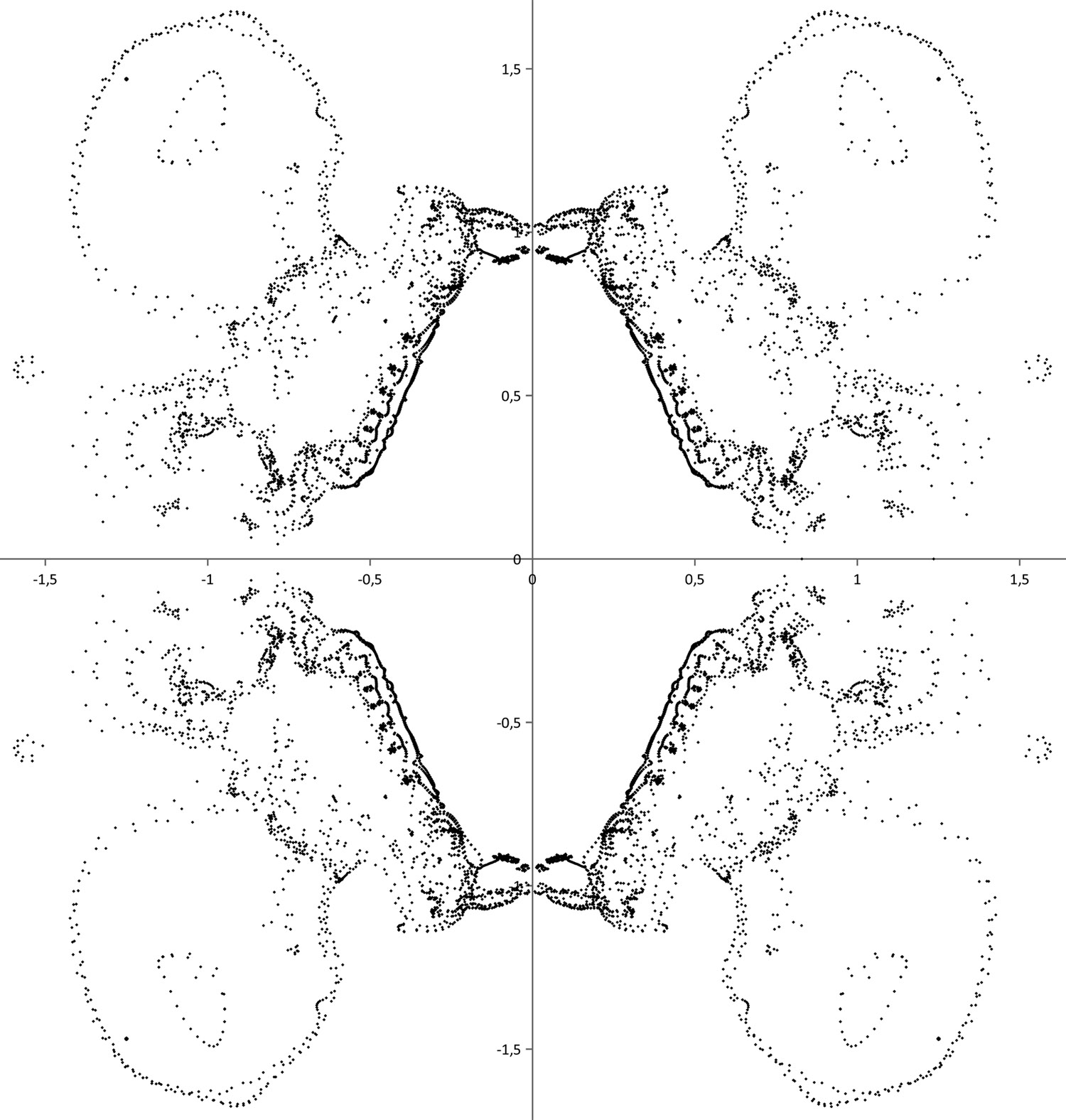}
\captionof{figure}{Spectrum of the Ising model on $SC(7,5,3)$.} 
\label{specn7p5}
\end{center}

\begin{center}
\includegraphics[width=\hsize]{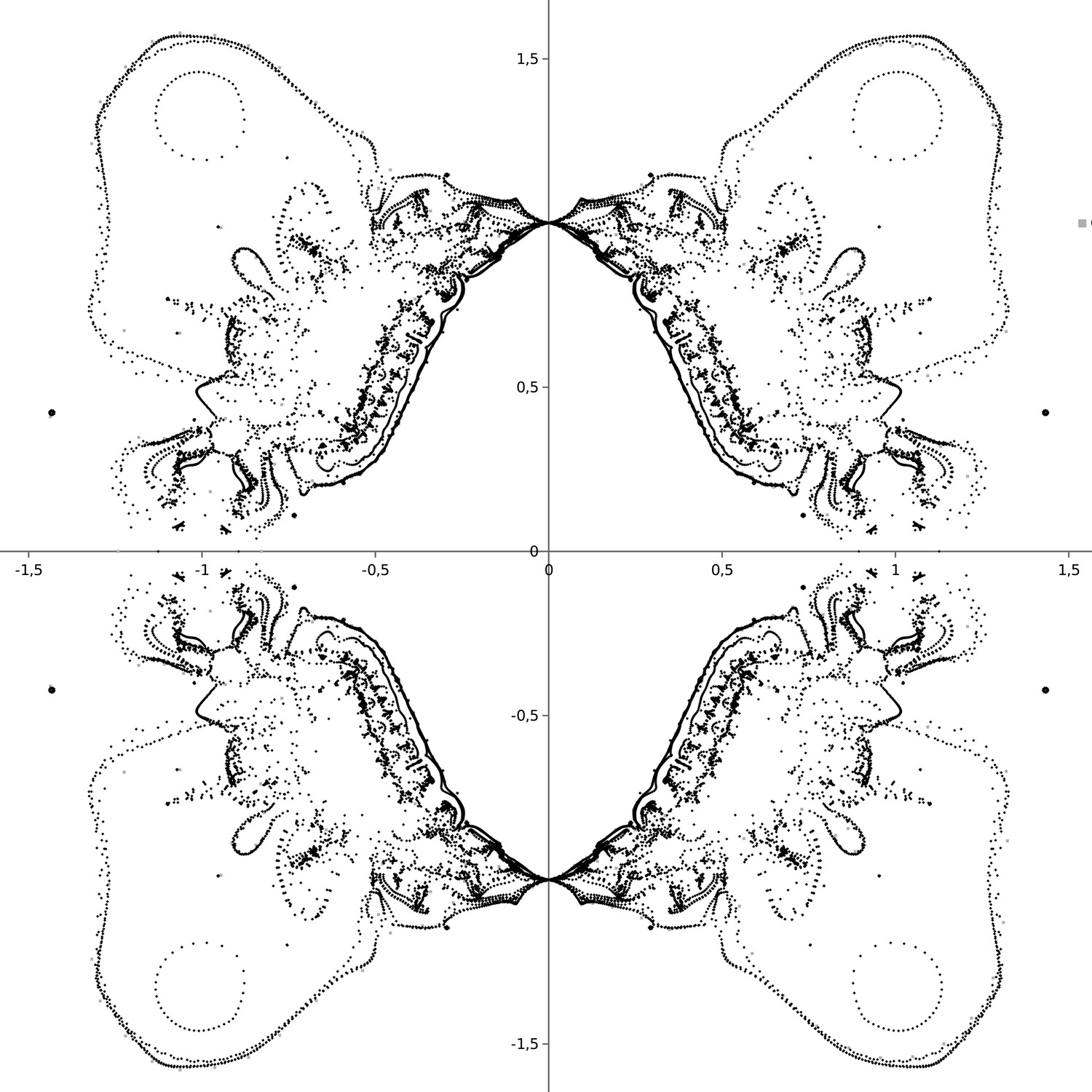}
\captionof{figure}{Spectrum of the Ising model on $SC(10,8,3)$.} 
\label{specn10p8k3}
\end{center}

\begin{center}
\includegraphics[width=\hsize]{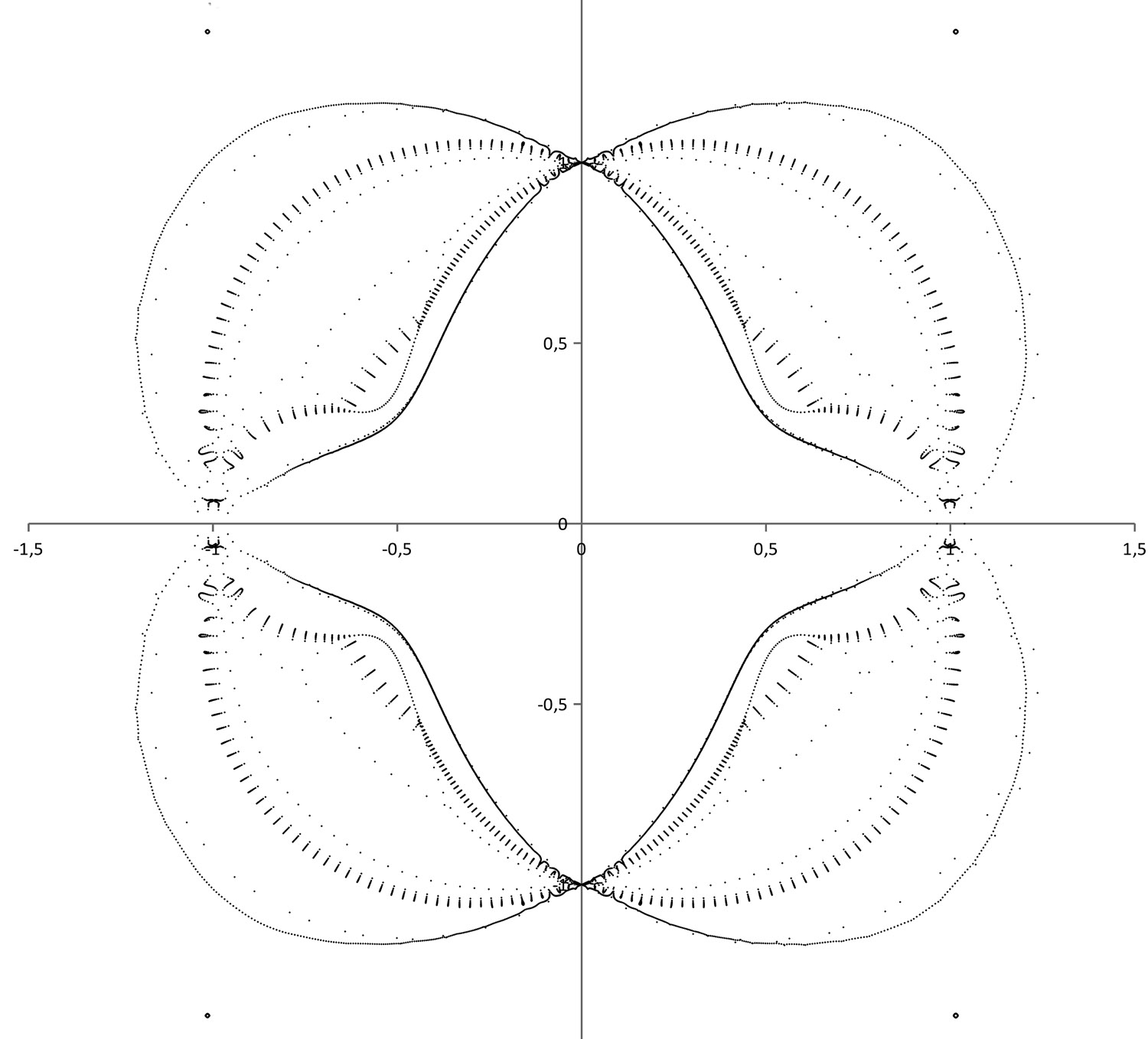}
\captionof{figure}{Spectrum of the Ising model on $SC(50,48,2)$.} 
\label{specn50p48k2}
\end{center}

\begin{center}
\includegraphics[width=\hsize]{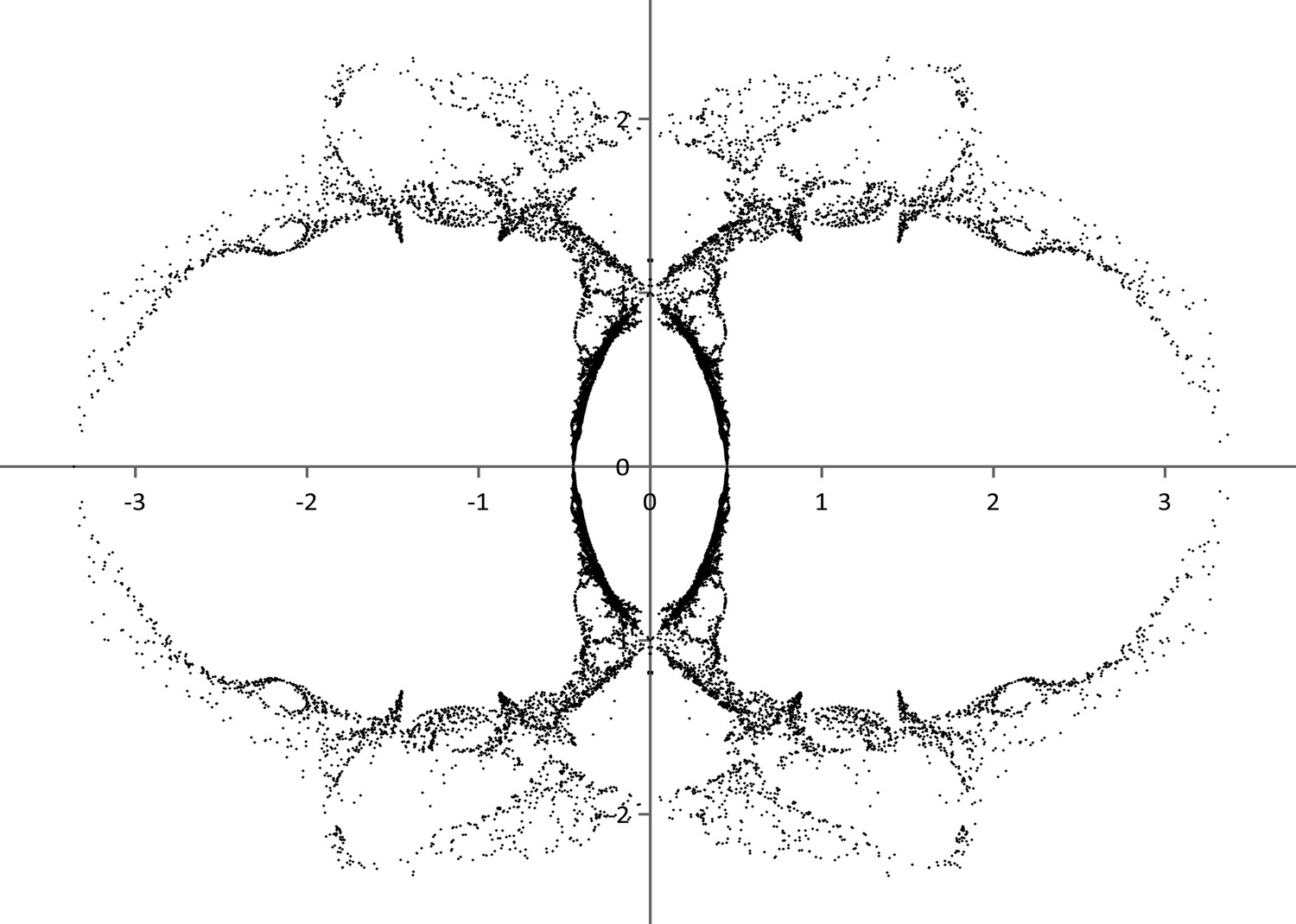}
\captionof{figure}{Spectrum of the Ising model on $SC(5,1,3)$.} 
\label{specn5p1k3}
\end{center}

\begin{center}
\includegraphics[width=\hsize]{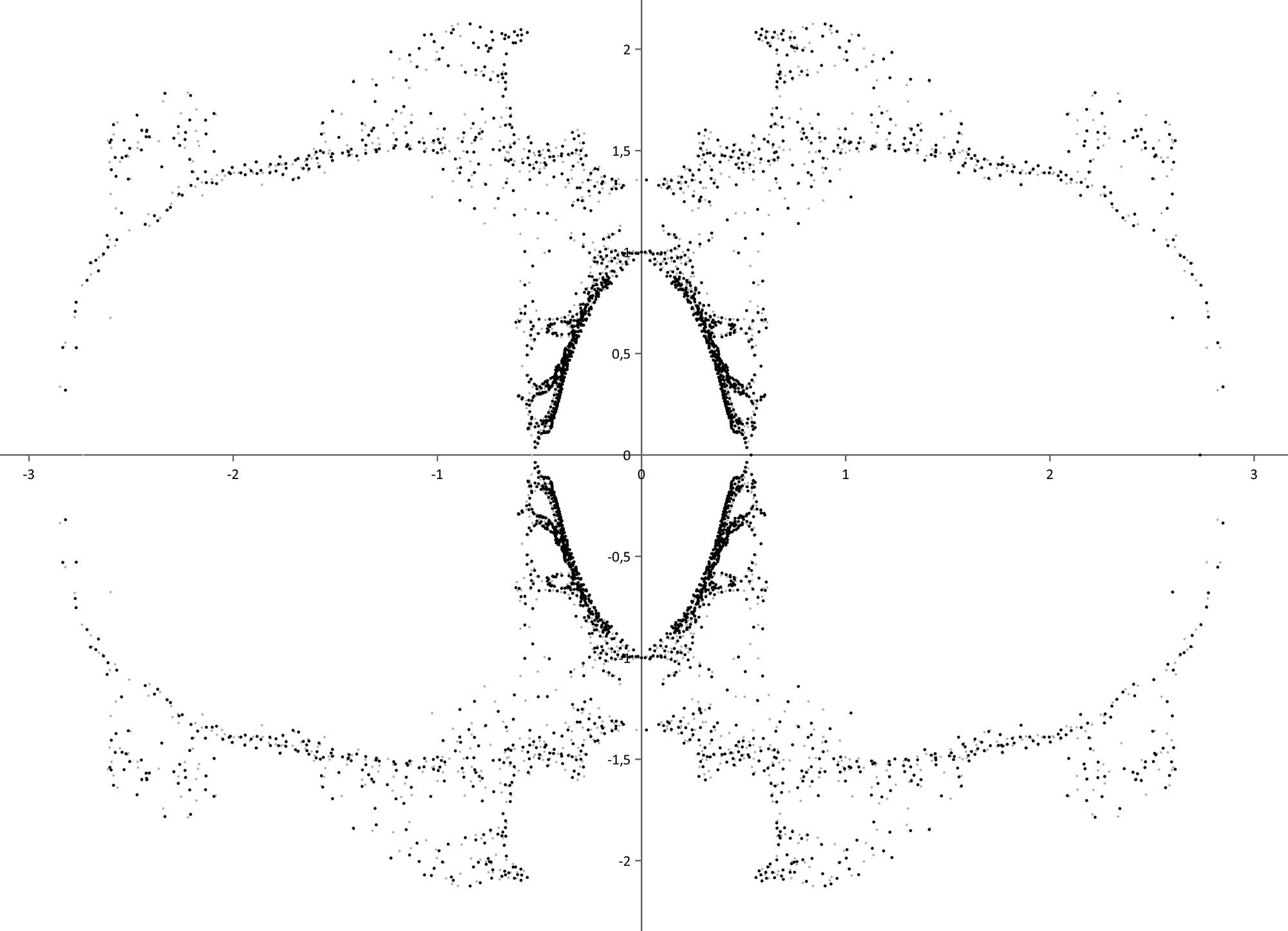}
\captionof{figure}{Spectrum of the Ising model on $SC(10,6,2)$.} 
\label{specn10p6k2}
\end{center}

\begin{center}
\includegraphics[width=\hsize]{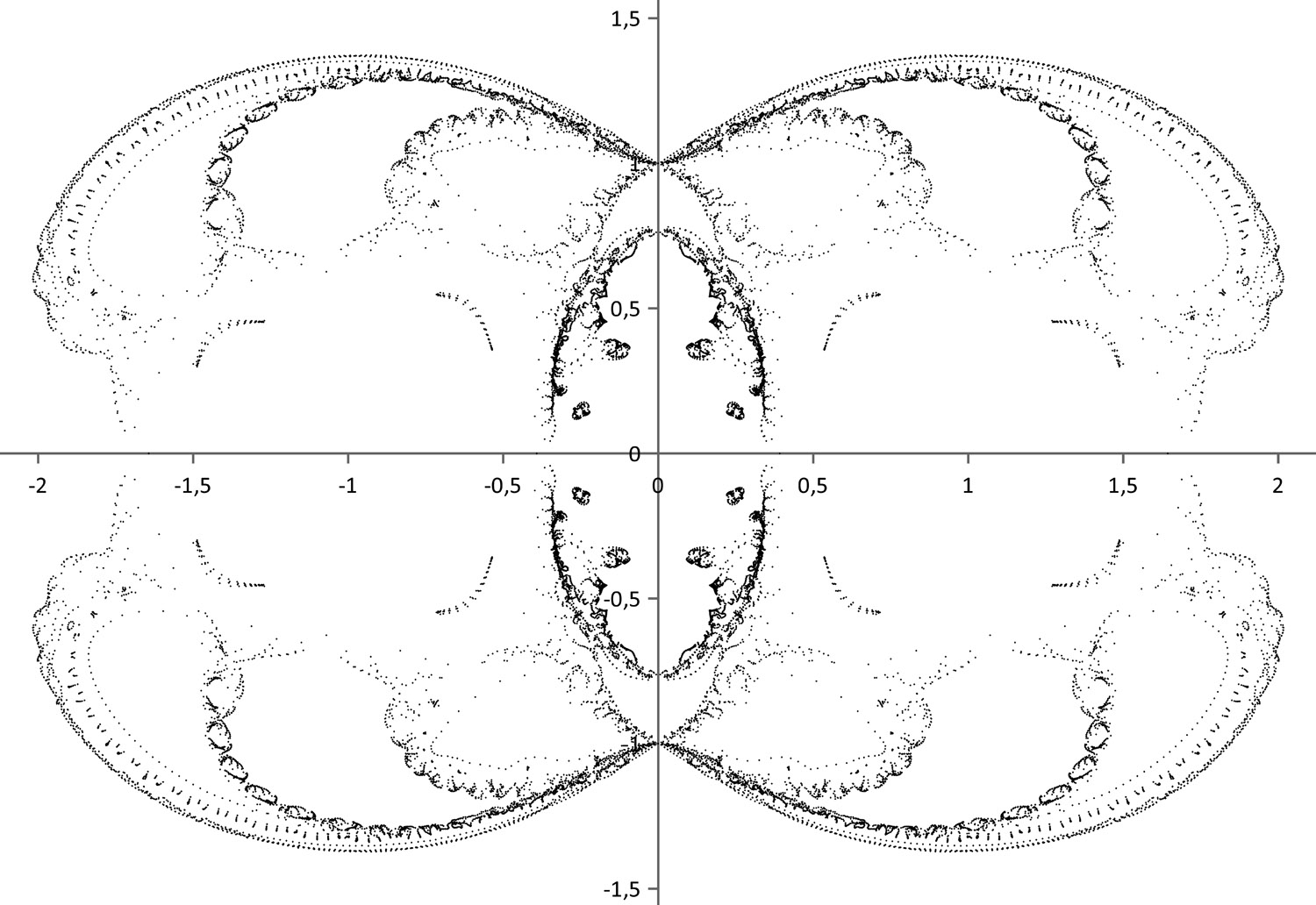}
\captionof{figure}{Spectrum of the Ising model on $SC(20,16,2)$.} 
\label{specn20p16k2}
\end{center}

\begin{center}
\includegraphics[width=\hsize]{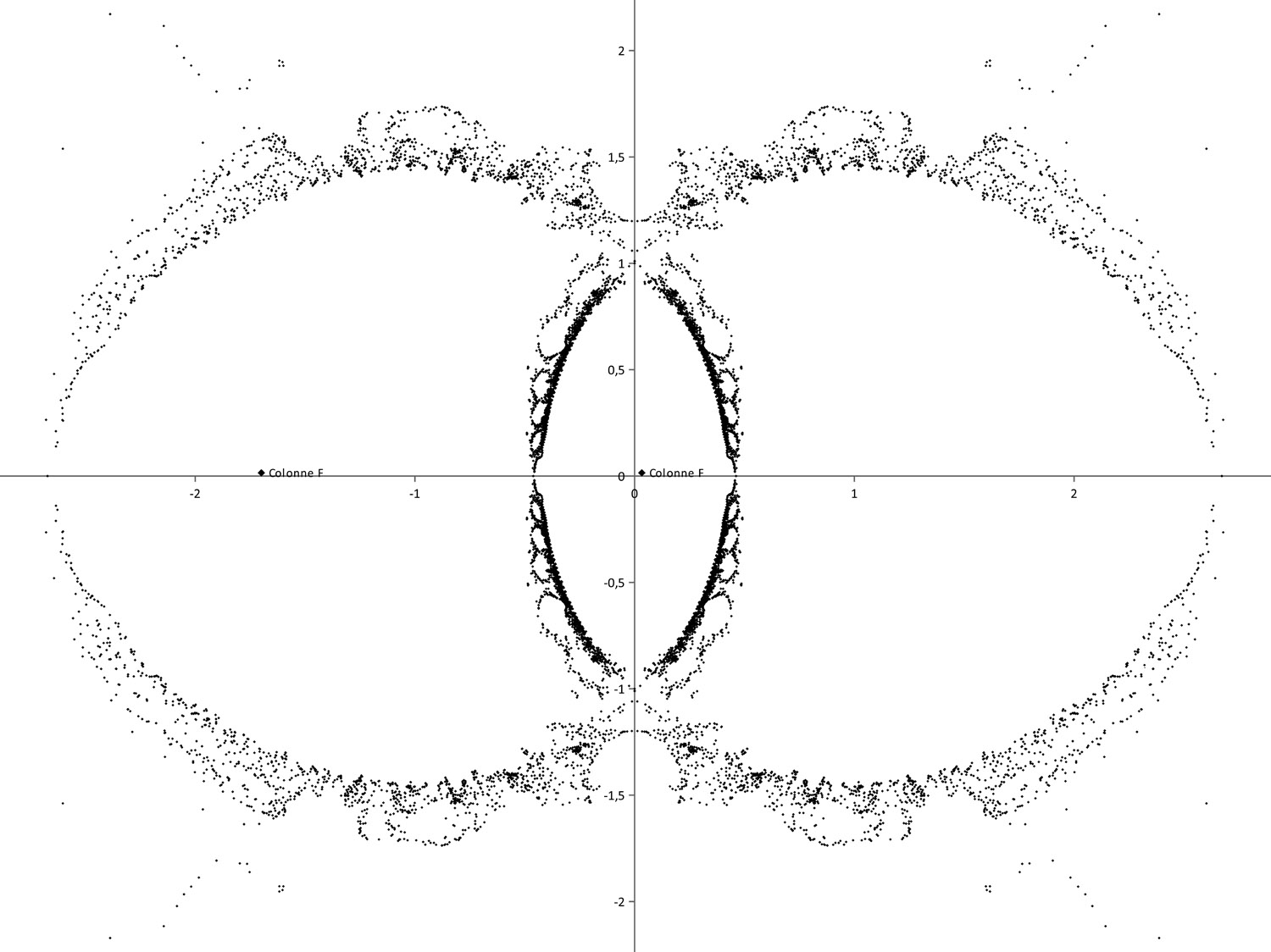}
\captionof{figure}{Spectrum of the Ising model on $SC(10,4,2)$.} 
\label{specn10p4k2}
\end{center}

\begin{center}
\includegraphics[width=\hsize]{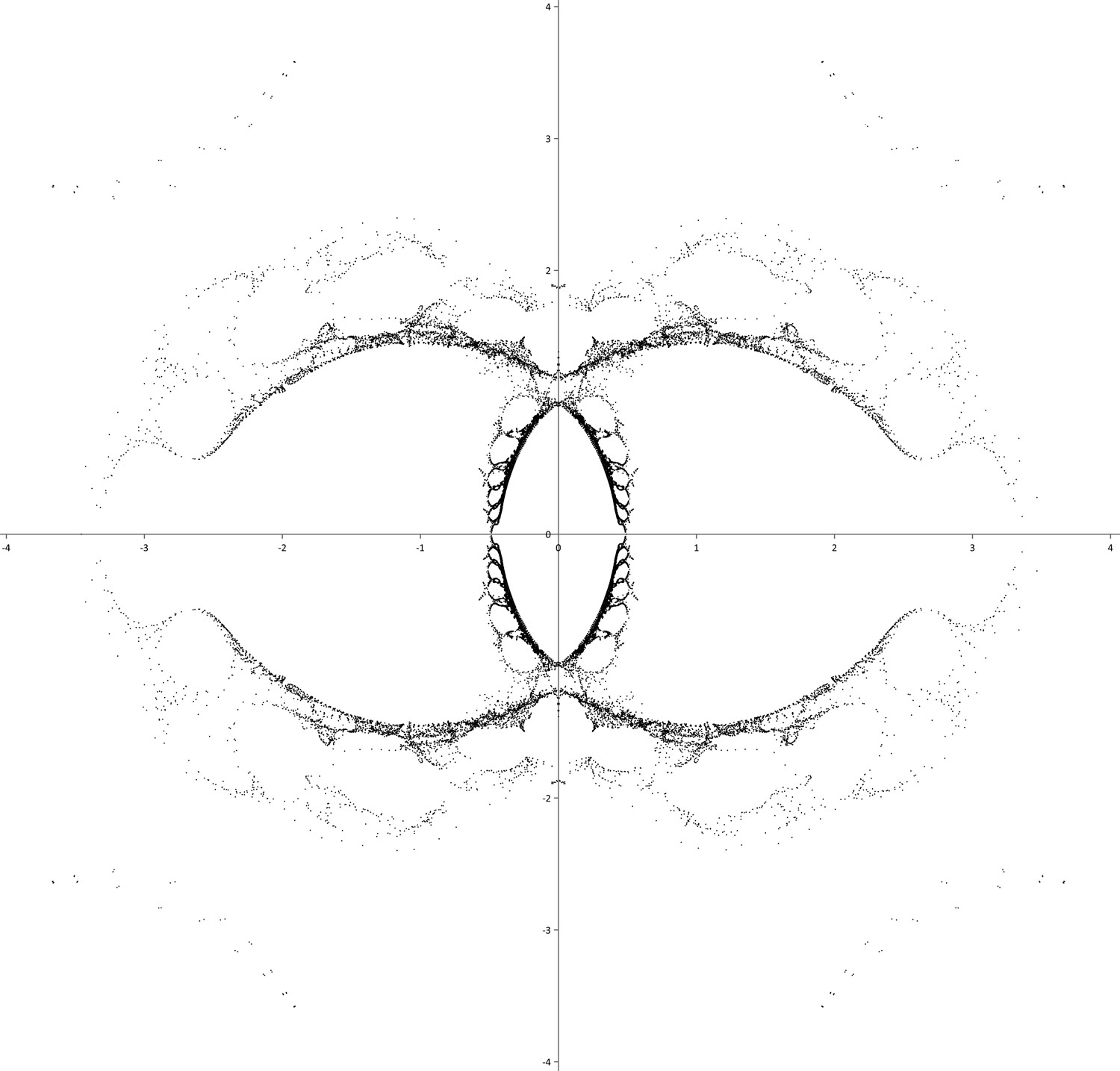}
\captionof{figure}{Spectrum of the Ising model on $SC(15,9,2)$.} 
\label{specn15p9k2}
\end{center}

\onecolumngrid
\section{The partition function of the Ising model on $SC(3,1,2)$}

$SC(3,1,2)$ contains $64$ sites $110$ bonds by pattern.

\begin{equation}
\begin{scriptsize}
\begin{aligned}
Z\ =&\ 2^{64N^2}\cosh(\beta J)^{110N^2}P_{N,SC(3,1)}(v)\ =\ 2^{\cal N}\ {P_{N,SC(3,1)}(v)\over (1-v^2)^{55N^2}}\\
=&\ {2^{\cal N}\over (1-v^2)^{55N^2}}\ {\prod_{p,q=1}^N}\left[R_1(v)+R_2(v)\left(\cos \left(\frac{8 p \pi}{n}\right)+\cos \left(\frac{8 \pi  q}{n}\right)\right)+R_3(v) \left(\cos \left(\frac{6 p \pi }{n}\right) \cos \left(\frac{2 \pi  q}{n}\right)+\cos \left(\frac{2 p \pi }{n}\right) \cos \left(\frac{6 \pi  q}{n}\right)\right)\right.\\
&\ +R_4(v) \cos \left(\frac{4 p \pi }{n}\right) \cos \left(\frac{4 \pi  q}{n}\right)+R_5(v) \left(\cos \left(\frac{6 p \pi }{n}\right)+\cos \left(\frac{6 \pi  q}{n}\right)\right)+R_6(v) \left(\cos \left(\frac{4 p \pi }{n}\right) \cos \left(\frac{2 \pi  q}{n}\right)+\cos \left(\frac{2 p \pi }{n}\right) \cos \left(\frac{4 \pi  q}{n}\right)\right)\\
&\ \left.\qquad+R_7(v)\left(\cos \left(\frac{4 p \pi }{n}\right)+\cos \left(\frac{4 \pi q}{n}\right)\right)+R_8(v) \cos \left(\frac{2 p \pi}{n}\right) \cos \left(\frac{2 \pi  q}{n}\right)+R_9(v)\left(\cos \left(\frac{2 p \pi }{n}\right)+\cos \left(\frac{2 \pi  q}{n}\right)\right)\right]^{1\over 2}
\end{aligned}
\end{scriptsize}
\label{partitionsca}
\end{equation}
where

$
\begin{scriptsize}
\setlength{\jot}{1pt}\setlength{\arraycolsep}{1pt}
\begin{array}{ll}
R_1(v)=&\left(v^2+1\right)^4 \left(66049 v^{160}+1863748 v^{158}+34816794 v^{156}+487920736 v^{154}+5829419081 v^{152}+62043615292 v^{150}+610375484810 v^{148}\right.\\
&+5629926712376 v^{146}+48958702058888 v^{144}+399190574073332 v^{142}+3020026447188996 v^{140}+20939175352666688 v^{138}\\
&+131593530887327595 v^{136}+742887817909110368 v^{134}+3741424964919899252 v^{132}+16726427461923894120 v^{130}\\
&+66164693351103870761 v^{128}+231231159684261562856 v^{126}+713993970461290556988 v^{124}+1950877964938315361368 v^{122}\\
&+4729945313928133002397 v^{120}+10214323037220920245824 v^{118}+19734846868999758671704 v^{116}+34280717346360845823824 v^{114}\\
&+53807872945283982704510 v^{112}+76699143848044407710980 v^{110}+99764536551012245567364 v^{108}+118955458176223051610336 v^{106}\\
&+130578198461606834048299 v^{104}+132483686836425068605656 v^{102}+124699322436786971114040 v^{100}\\
&+109261488136286331184772 v^{98}+89404899139214195794562 v^{96}+68524748500512477074664 v^{94}+49334331878343868910400 v^{92}\\
&+33452158116700655711816 v^{90}+21417630707307219041287 v^{88}+12978900917016138639416 v^{86}+7461425249091033310236 v^{84}\\
&+4078268187776828657176 v^{82}+2123783358962834015018 v^{80}+1055834356380309541172 v^{78}+502071810081383801644 v^{76}\\
&+228780112298202683664 v^{74}+100073283521221762961 v^{72}+42092218247019149464 v^{70}+17052314233752687552 v^{68}\\
&+6664449590236547004 v^{66}+2516787272463476670 v^{64}+919917855909227880 v^{62}+326006280619324900 v^{60}+112223062195620584 v^{58}\\
&+37600779228535879 v^{56}+12288317194266400 v^{54}+3926744284883672 v^{52}+1229476141619136 v^{50}+378398010013938 v^{48}\\
&+114536990090236 v^{46}+34304590838540 v^{44}+10099319972160 v^{42}+2978033108769 v^{40}+850312388104 v^{38}+251460219976 v^{36}\\
&+67929819868 v^{34}+20962865237 v^{32}+4981759236 v^{30}+1749288582 v^{28}+309457944 v^{26}+146666828 v^{24}+13210924 v^{22}\\
&\left.+12043562 v^{20}+14880 v^{18}+906014 v^{16}-55720 v^{14}+56640 v^{12}-4976 v^{10}+2616 v^8-216 v^6+76 v^4-4 v^2+1\right)\\
\end{array}
\end{scriptsize}
$
\vskip2mm
$
\begin{scriptsize}
\setlength{\jot}{1pt}\setlength{\arraycolsep}{1pt}
\begin{array}{ll}
R_2(v)=&-2 v^{36}\left(v^2-1\right)^{25} \left(v^2+1\right)^4 \left(3 v^4+4 v^2+1\right)^5 \left(12 v^{18}+162 v^{16}+381 v^{14}+377 v^{12}+423 v^{10}+353 v^8+235 v^6+83 v^4+21 v^2+1\right)\\
&\qquad \left(4 v^{22}+138 v^{20}+725 v^{18}+2043 v^{16}+2032 v^{14}+1526 v^{12}+1054 v^{10}+412 v^8+204 v^6+40 v^4+13 v^2+1\right)\\
\end{array}
\end{scriptsize}
$
\vskip2mm
$
\begin{scriptsize}
\setlength{\jot}{1pt}\setlength{\arraycolsep}{1pt}
\begin{array}{ll}
R_3(v)=&-4 v^{36}\left(v^2-1\right)^{22}\left(v^2+1\right)^7 \left(3 v^2+1\right)^2 \left(1552 v^{60}+104660 v^{58}+2623048 v^{56}+36544175 v^{54}+330270281 v^{52}+2101336447 v^{50}\right.\\
&+9885230641 v^{48}+35504293289 v^{46}+100142127302 v^{44}+227642548893 v^{42}+426361617874 v^{40}+670063319582 v^{38}\\
&+896919150247 v^{36}+1034933065502 v^{34}+1039313018603 v^{32}+915147933074 v^{30}+710561230372 v^{28}+488475359822 v^{26}\\
&+298100123604 v^{24}+161696422331 v^{22}+77940138263 v^{20}+33322715115 v^{18}+12589249735 v^{16}+4177137653 v^{14}+1206199750 v^{12}\\
&+299106289 v^{10}+62475282 v^8+10670664 v^6\left.+1423161 v^4+135624 v^2+7709\right)\\
\end{array}
\end{scriptsize}
$
\vskip2mm
$
\begin{scriptsize}
\setlength{\jot}{1pt}\setlength{\arraycolsep}{1pt}
\begin{array}{ll}
R_4(v)=&4 v^{36} \left(v^2-1\right)^{20} \left(v^2+1\right)^7 \left(16 v^{74}+1136 v^{72}+44808 v^{70}+1113848 v^{68}+18788249 v^{66}+226745511 v^{64}+2056866845 v^{62}\right.\\
&+14605056521 v^{60}+83422751764 v^{58}+389934141928 v^{56}+1506551494055 v^{54}+4832755748853 v^{52}+12930769847443 v^{50}\\
&+29083407454137 v^{48}+55531739530829 v^{46}+90949872937197 v^{44}+129032770346814 v^{42}+159984513481478 v^{40}\\
&+174697858136455 v^{38}+169116989529009 v^{36}+145938181676275 v^{34}+112768244273345 v^{32}+78303840103391 v^{30}\\
&+48989971397019 v^{28}+27664691953048 v^{26}+14113232052108 v^{24}+6504372564525 v^{22}+2705489437327 v^{20}\\
&+1013571431133 v^{18}+340849058231 v^{16}+102355080055 v^{14}+27236070639 v^{12}+6347636774 v^{10}\\
&\left.+1273210614 v^8+213673437 v^6+28625235 v^4+2781924 v^2+155976\right)\\
\end{array}
\end{scriptsize}
$
\vskip2mm
$
\begin{scriptsize}
\setlength{\jot}{1pt}\setlength{\arraycolsep}{1pt}
\begin{array}{ll}
R_5(v)=&-4 v^{27}\left(v^2-1\right)^{16}\left(v^2+1\right)^8 \left(3 v^2+1\right)^2 \left(2 v^4+v^2+1\right) \left(1728 v^{76}+116180 v^{74}+3178068 v^{72}+49686415 v^{70}+517894437 v^{68}\right.\\
&+3886572161 v^{66}+21884115762 v^{64}+94798014997 v^{62}+320997452955 v^{60}+862911495039 v^{58}+1877655396613 v^{56}\\
&+3389560358223 v^{54}+5216261979834 v^{52}+7008863091397 v^{50}+8362045367138 v^{48}+8958623022805 v^{46}+8685449796189 v^{44}\\
&+7664188160591 v^{42}+6180269415883 v^{40}+4566174560301 v^{38}+3094910703210 v^{36}+1924846945407 v^{34}+1097676431528 v^{32}\\
&+573205607463 v^{30}+273548809509 v^{28}+119009889125 v^{26}+47054202075 v^{24}+16841739821 v^{22}+5430679242 v^{20}+1567289311 v^{18}\\
&\left. +401399074 v^{16}+90134895 v^{14}+17430275 v^{12}+2832121 v^{10}+369057 v^8+36424 v^6+2317 v^4+92 v^2+2\right)\\
\end{array}
\end{scriptsize}
$
\vskip2mm
$
\begin{scriptsize}
\setlength{\jot}{1pt}\setlength{\arraycolsep}{1pt}
\begin{array}{ll}
R_6(v)=&8 v^{27}\left(v^2-1\right)^{14} \left(v^2+1\right)^6 \left(2 v^4+v^2+1\right) \left(64 v^{94}+4468 v^{92}+147980 v^{90}+3179693 v^{88}+48100343 v^{86}+539997553 v^{84}\right.\\
&+4770395183 v^{82}+35083500198 v^{80}+224191379662 v^{78}+1270897628916 v^{76}+6353979262310 v^{74}+27419894552939 v^{72}\\
&+99692249388051 v^{70}+299571723554576 v^{68}+734558942945704 v^{66}+1455629494837068 v^{64}+2298609868769048 v^{62}\\
&+2789873400545836 v^{60}+2294821524686598 v^{58}+408197671747208 v^{56}-2680408652915190 v^{54}-6162947800014706 v^{52}\\
&-8990178899390348 v^{50}-10388761517933962 v^{48}-10175490371357988 v^{46}-8719086691451452 v^{44}-6656329708197718 v^{42}\\
&-4581651893393268 v^{40}-2867329175807710 v^{38}-1641666837075932 v^{36}-863901046287898 v^{34}-419305659241182 v^{32}\\
&-188185970317048 v^{30}-78228572210592 v^{28}-30147290214682 v^{26}-10771291722461 v^{24}-3565251287361 v^{22}\\
&-1091357780071 v^{20}-308085312147 v^{18}-79866998132 v^{16}-18899254154 v^{14}-4047913912 v^{12}-775289992 v^{10}-130506063 v^8\\
&\left. -18777717 v^6-2212812 v^4-194334 v^2-10518\right)\\
\end{array}
\end{scriptsize}
$
\vskip2mm
$
\begin{scriptsize}
\setlength{\jot}{1pt}\setlength{\arraycolsep}{1pt}
\begin{array}{ll}
R_7(v)=&2 v^{18}\left(v^2-1\right)^{10} \left(v^2+1\right)^5 \left(1024 v^{118}+73376 v^{116}+2554100 v^{114}+61680272 v^{112}+1135766637 v^{110}+16345369066 v^{108}\right.\\
&+187411659025 v^{106}+1741846959450 v^{104}+13320720942411 v^{102}+84900402219031 v^{100}+456067876580770 v^{98}\\
&+2086336712639254 v^{96}+8207499502923847 v^{94}+28018244480392268 v^{92}+83649885538066464 v^{90}+219758953768469198 v^{88}\\
&+510345430344449083 v^{86}+1051460444511168279 v^{84}+1928724075301782099 v^{82}+3162816604789189952 v^{80}\\
&+4659238048491544778 v^{78}+6199531863618509022 v^{76}+7493232156162353593 v^{74}+8272752541794572812 v^{72}\\
&+8385809981467811224 v^{70}+7841544649922062640 v^{68}+6793384811200483187 v^{66}+5474236572130919434 v^{64}\\
&+4118482269989569790 v^{62}+2903189453778909336 v^{60}+1924043621020269368 v^{58}+1202659066830751772 v^{56}\\
&+711079447233785030 v^{54}+398683945429932494 v^{52}+212394624253965554 v^{50}+107666539529239952 v^{48}\\
&+51975085641684897 v^{46}+23899531281956182 v^{44}+10465128030212419 v^{42}+4360767896957906 v^{40}+1727465990013995 v^{38}\\
&+649748893940119 v^{36}+231712754730552 v^{34}+78221937383930 v^{32}+24952168980907 v^{30}+7506287117724 v^{28}\\
&+2124645600872 v^{26}+564354657718 v^{24}+140226063375 v^{22}+32471209675 v^{20}+6972361047 v^{18}\\
&\left.+1380398976 v^{16}+249713208 v^{14}+40880994 v^{12}+5940275 v^{10}+751144 v^8+78306 v^6+6354 v^4+291 v^2+6\right)\\
\end{array}
\end{scriptsize}
$
\vskip2mm
$
\begin{scriptsize}
\setlength{\jot}{1pt}\setlength{\arraycolsep}{1pt}
\begin{array}{ll}
R_8(v)=&-4 v^{18}\left(v^2-1\right)^8 \left(v^2+1\right)^5 \left(8 v^{122}-43744 v^{120}+434415 v^{118}+56685227 v^{116}+1582898395 v^{114}+26973878029 v^{112}\right.\\
&+332232767684 v^{110}+3161647080670 v^{108}+24022764747379 v^{106}+148113154974329 v^{104}+745063037410971 v^{102}\\
&+3044426399657783 v^{100}+9929123575565031 v^{98}+24717303735125867 v^{96}+41006099138180148 v^{94}+14943288851876238 v^{92}\\
&-170199034212356451 v^{90}-688468364610506627 v^{88}-1611489604584985055 v^{86}-2590190737268385047 v^{84}\\
&-2578322417918731080 v^{82}+41886337410603478 v^{80}+6738996949380983332 v^{78}+17810610926201164400 v^{76}\\
&+31746632728603606674 v^{74}+45527702609694782768 v^{72}+55788871200943083381 v^{70}+60185308971036409745 v^{68}\\
&+58203935163856867590 v^{66}+51084600818381737534 v^{64}+41064111709601730136 v^{62}+30447946031002699964 v^{60}\\
&+20944918133200798994 v^{58}+13430943889071569690 v^{56}+8061353392932011321 v^{54}+4544664561061130837 v^{52}\\
&+2413828463455857579 v^{50}+1211079701115028793 v^{48}+575319870218788828 v^{46}+259298889151315102 v^{44}\\
&+111075658093223063 v^{42}+45293042975989945 v^{40}+17603611271047901 v^{38}+6528087293378881 v^{36}\\
&+2311645362984171 v^{34}+782016112574191 v^{32}+252765221818644 v^{30}+78038916297686 v^{28}+22995578384473 v^{26}\\
&+6459282938969 v^{24}+1725823202295 v^{22}+437578372535 v^{20}+104829369618 v^{18}+23649312116 v^{16}+4980172188 v^{14}\\
&\left.+976168436 v^{12}+174407252 v^{10}+28562062 v^8+4025043 v^6+514647 v^4+46296 v^2+4056\right)\\
\end{array}
\end{scriptsize}
$
\vskip2mm
$
\begin{scriptsize}
\setlength{\jot}{1pt}\setlength{\arraycolsep}{1pt}
\begin{array}{ll}
R_9(v)=&-4 v^9\left(v^2-1\right)^4 \left(v^2+1\right)^4 \left(2 v^4+v^2+1\right) \left(4112 v^{138}+200909 v^{136}+5279060 v^{134}+102919848 v^{132}+1592164454 v^{130}\right.\\
&+20374170747 v^{128}+220568043592 v^{126}+2055887653156 v^{124}+16701392030854 v^{122}+119261415980801 v^{120}\\
&+752450783040696 v^{118}+4206423069829700 v^{116}+20858948937177312 v^{114}+91750445494799624 v^{112}\\
&+357722842944128874 v^{110}+1235144379705618348 v^{108}+3774672373739640158 v^{106}+10211734492259744752 v^{104}\\
&+24481670689937587754 v^{102}+52119197696490061447 v^{100}+98831597011702566762 v^{98}+167592655387320913669 v^{96}\\
&+255337783769321807924 v^{94}+351356860257986647848 v^{92}+439111867018551017767 v^{90}+501288088633013066403 v^{88}\\
&+525749750441301135784 v^{86}+509422440738518382176 v^{84}+458437420569107511257 v^{82}+385026430220859258864 v^{80}\\
&+303098245415815930870 v^{78}+224482491373995456348 v^{76}+156915478237408921061 v^{74}+103797782877985781292 v^{72}\\
&+65120042524870903034 v^{70}+38820595387615963383 v^{68}+22026283694224307181 v^{66}+11912301786843929535 v^{64}\\
&+6149415270337165824 v^{62}+3034232762893964580 v^{60}+1432945264222755180 v^{58}+648578400898673015 v^{56}\\
&+281726091012094072 v^{54}+117594256081173484 v^{52}+47224835997992010 v^{50}+18266898286316320 v^{48}\\
&+6812217382030742 v^{46}+2451260713877396 v^{44}+851574232585624 v^{42}+285739596480100 v^{40}+92619873272814 v^{38}\\
&+29004260127893 v^{36}+8772850615388 v^{34}+2562514352671 v^{32}+722406708100 v^{30}+196421464112 v^{28}+51478939607 v^{26}\\
&+12967023589 v^{24}+3144888808 v^{22}+725803792 v^{20}+161734317 v^{18}+33378952 v^{16}+6775050 v^{14}\\
&\left.+1189828 v^{12}+218149 v^{10}+29619 v^8+4826 v^6+421 v^4+55 v^2+2\right)\\
\end{array}
\end{scriptsize}
$
\vskip2mm
so that
\vskip2mm
$
\begin{scriptsize}
\setlength{\jot}{0pt}\setlength{\arraycolsep}{1pt}
\begin{array}{ll}
P_{1,SC(3,1,2)}(v)=&\left(v^2+1\right)^2 \left(257 v^{80}-128 v^{79}+3602 v^{78}-4008 v^{77}+41339 v^{76}-66020 v^{75}+314996 v^{74}-923204v^{73}+2022356 v^{72}\right.\\
&-9483940 v^{71}+9962448 v^{70}-74660868 v^{69}+37916277 v^{68}-453962036 v^{67}+94318076v^{66}-2170108652 v^{65}+148626139 v^{64}\\
&-8091996788 v^{63}+52373854 v^{62}-23345186256 v^{61}-316690249 v^{60}-52396150344v^{59}-666668564 v^{58}-93531471156 v^{57}\\
&-410332080 v^{56}-136245425748 v^{55}+310209308 v^{54}-165405606644v^{53}+783915541 v^{52}-170322391948 v^{51}+595813070 v^{50}\\
&-151098217012 v^{49}+58095160 v^{48}-117028842596v^{47}-266939254 v^{46}-80080515568 v^{45}-265452631 v^{44}-48879509912 v^{43}\\
&-162835668 v^{42}-26841909236 v^{41}-84027328v^{40}-13360975020 v^{39}-29486812 v^{38}-6081169068 v^{37}+9235899 v^{36}\\
&-2551455372 v^{35}+26920766 v^{34}-996250996v^{33}+28648236 v^{32}-363038748 v^{31}+20224474 v^{30}-124857072 v^{29}\\
&+12751389 v^{28}-40338664 v^{27}+6192644v^{26}-12348684 v^{25}+3081424 v^{24}-3544924 v^{23}+1085332 v^{22}-926556 v^{21}\\
&+489559 v^{20}-232756 v^{19}+116346v^{18}-46316 v^{17}+60223 v^{16}-10124 v^{15}+6828 v^{14}-1192 v^{13}+6472 v^{12}-212 v^{11}\\
&\left.-16 v^{10}-8 v^9+588 v^8-36v^6+36 v^4-2 v^2+1\right)
\end{array}
\end{scriptsize}
$
\vglue2mm
{\parindent0pt whose real root between $0$ and $1$ is $v_c\approx 0.5815$ corresponding to the critical temperature $T_c\approx 1.654$. $v_c$ is also the single real root between $0$ and $1$ of}
\vskip2mm
$
\begin{scriptsize}
\setlength{\jot}{1pt}\setlength{\arraycolsep}{1pt}
\begin{array}{ll}
Q^-_{1,SC(3,1,2)}(v)=&\left(v^2+1\right) \left(\left(17-4 \sqrt{2}\right) v^{40}+\left(16-42\sqrt{2}\right) v^{39}+\left(229-57 \sqrt{2}\right)v^{38}+\left(136-399 \sqrt{2}\right) v^{37}+\left(1982-582\sqrt{2}\right) v^{36}\right.\\
&+\left(1388-3136 \sqrt{2}\right)v^{35}+\left(11593-4493 \sqrt{2}\right) v^{34}+\left(4378-13696\sqrt{2}\right) v^{33}+\left(47345-20943 \sqrt{2}\right)v^{32}\\
&+\left(4390-43509 \sqrt{2}\right) v^{31}+\left(120355-59631\sqrt{2}\right) v^{30}+2 \left(-524-40439 \sqrt{2}\right)v^{29}+\left(193694-101453 \sqrt{2}\right) v^{28}\\
&+2\left(-4115-52474 \sqrt{2}\right) v^{27}+\left(229483-114973\sqrt{2}\right) v^{26}+9 \left(-808-11583 \sqrt{2}\right)v^{25}+\left(194468-100371 \sqrt{2}\right) v^{24}\\
&+\left(-1760-79069\sqrt{2}\right) v^{23}+\left(127651-65193 \sqrt{2}\right)v^{22}+\left(1298-50560 \sqrt{2}\right) v^{21}+\left(69620-33493\sqrt{2}\right) v^{20}\\
&+48 \left(58-535 \sqrt{2}\right)v^{19}+\left(32407-14607 \sqrt{2}\right) v^{18}+\left(1786-11675\sqrt{2}\right) v^{17}+\left(13306-5693 \sqrt{2}\right)v^{16}+\left(1222-4263 \sqrt{2}\right) v^{15}\\
&+\left(4261-2029\sqrt{2}\right) v^{14}+\left(668-1522 \sqrt{2}\right)v^{13}+\left(1598-599 \sqrt{2}\right) v^{12}+\left(186-444\sqrt{2}\right) v^{11}+\left(341-135 \sqrt{2}\right)v^{10}\\
&+\left(52-173 \sqrt{2}\right) v^9+\left(191-29 \sqrt{2}\right)v^8+\left(4-29 \sqrt{2}\right) v^7\left.+2 \left(8-\sqrt{2}\right)v^6+\left(2-17 \sqrt{2}\right) v^5+\left(18-\sqrt{2}\right)v^4-\sqrt{2} v+1\right)
\end{array}
\end{scriptsize}
$
\vskip2mm
The polynom $Q^+_{1,SC(3,1,2)}(v)$ is obtained by the substitution $-\sqrt{2}\rightarrow +\sqrt{2}$ in $Q^-_{1,SC(3,1,2)}(v)$

\section{The partition function of the Ising model on $SM(3,2)$}

$SM(3,2)$ contains $60$ sites and $90$ bonds by pattern.

\begin{equation}
\begin{scriptsize}
\begin{aligned}
Z\ =&\ 2^{60N^2}\cosh(\beta J)^{90N^2}P_{N,SC(3,1)}(v)\ =\ 2^{\cal N}\ {P_{N,SC(3,1)}(v)\over (1-v^2)^{45N^2}}\\
=&\ {2^{\cal N}\over (1-v^2)^{45N^2}}\ {\prod_{p,q=1}^N}\left[R_1(v)+R_2(v)\left(\cos \left(\frac{8 p \pi}{n}\right)+\cos \left(\frac{8 \pi  q}{n}\right)\right)+R_3(v) \left(\cos \left(\frac{6 p \pi }{n}\right) \cos \left(\frac{2 \pi  q}{n}\right)+\cos \left(\frac{2 p \pi }{n}\right) \cos \left(\frac{6 \pi  q}{n}\right)\right)\right.\\
&\ +R_4(v) \cos \left(\frac{4 p \pi }{n}\right) \cos \left(\frac{4 \pi  q}{n}\right)+R_5(v) \left(\cos \left(\frac{6 p \pi }{n}\right)+\cos \left(\frac{6 \pi  q}{n}\right)\right)+R_6(v) \left(\cos \left(\frac{4 p \pi }{n}\right) \cos \left(\frac{2 \pi  q}{n}\right)+\cos \left(\frac{2 p \pi }{n}\right) \cos \left(\frac{4 \pi  q}{n}\right)\right)\\
&\ \left.\qquad+R_7(v)\left(\cos \left(\frac{4 p \pi }{n}\right)+\cos \left(\frac{4 \pi q}{n}\right)\right)+R_8(v) \cos \left(\frac{2 p \pi}{n}\right) \cos \left(\frac{2 \pi  q}{n}\right)+R_9(v)\left(\cos \left(\frac{2 p \pi }{n}\right)+\cos \left(\frac{2 \pi  q}{n}\right)\right)\right]^{1\over 2}
\end{aligned}
\end{scriptsize}
\label{partitionscb}
\end{equation}
where

$
\begin{scriptsize}
\setlength{\jot}{1pt}\setlength{\arraycolsep}{1pt}
\begin{array}{ll}
R_1(v)=&\left(v^2+1\right)^2 \left(4 v^{144}+168 v^{142}+3493 v^{140}+47698 v^{138}+502193 v^{136}+4548344 v^{134}+37644100 v^{132}+289220492 v^{130}+2052061788 v^{128}\right.\\
&+13203840312 v^{126}+75418854490 v^{124}+377705886692 v^{122}+1656678850504 v^{120}+6389178374964 v^{118}+21785292369054 v^{116}\\
&+66061262622480 v^{114}+179178091355156 v^{112}+437232338004104 v^{110}+965609067486671 v^{108}+1941179792606042 v^{106}\\
&+3571021346529995 v^{104}+6038485763430800 v^{102}+9420885932583585 v^{100}+13602852156396606 v^{98}+18227129771588581 v^{96}\\
&+22717052533473664 v^{94}+26390936136502940 v^{92}+28632226464394728 v^{90}+29069208869383966 v^{88}+27673689237383148 v^{86}\\
&+24759917135186564 v^{84}+20867009071686028 v^{82}+16606069264330538 v^{80}+12507528771092052 v^{78}+8937835724693451 v^{76}\\
&+6072987467598966 v^{74}+3932528520012443 v^{72}+2431554661398644 v^{70}+1438561668755527 v^{68}+815660979136414 v^{66}\\
&+444009273799171 v^{64}+232327270803528 v^{62}+117018251385354 v^{60}+56779716559572 v^{58}+26576679786140 v^{56}\\
&+12006589506428 v^{54}+5243985647630 v^{52}+2215681614536 v^{50}+908629321760 v^{48}+361998884608 v^{46}+140929025593 v^{44}\\
&+53588783614 v^{42}+20102829429 v^{40}+7385015048 v^{38}+2697334895 v^{36}+959189498 v^{34}+342267259 v^{32}+117727840 v^{30}\\
&+41556416 v^{28}+13673280 v^{26}+4923262 v^{24}+1454468 v^{22}+576016 v^{20}+127804 v^{18}+66270 v^{16}+7196 v^{14}\\
&\left.+7112 v^{12}+628 v^8-36 v^6+37 v^4-2 v^2+1\right)
\end{array}
\end{scriptsize}
$
\vskip2mm
$
\begin{scriptsize}
\setlength{\jot}{1pt}\setlength{\arraycolsep}{1pt}
\begin{array}{ll}
R_2(v)=&-2 (v-1)^{23} v^{36} (v+1)^{23} \left(v^2+1\right)^8 \left(11 v^8+27 v^6+18 v^4+7 v^2+1\right) \left(4 v^{26}+44 v^{24}+186 v^{22}+363 v^{20}+383 v^{18}+593 v^{16}\right.\\
&\left.+742 v^{14}+782 v^{12}+544 v^{10}+298 v^8+120 v^6+31 v^4+5 v^2+1\right)	
\end{array}
\end{scriptsize}
$
\vskip2mm
$
\begin{scriptsize}
\setlength{\jot}{1pt}\setlength{\arraycolsep}{1pt}
\begin{array}{ll}
R_3(v)=&-4 v^{36} \left(1-v^2\right)^{18} \left(v^2+1\right)^8 \left(27 v^{48}+1073 v^{46}+18487 v^{44}+186594 v^{42}+1245343 v^{40}+5862890 v^{38}+20289826 v^{36}\right.\\
&+53226652 v^{34}+108727643 v^{32}+177521314 v^{30}+237487670 v^{28}+266065120 v^{26}+254213358 v^{24}+209900620 v^{22}+151210772 v^{20}\\
&\left.+95620156 v^{18}+53197121 v^{16}+26011013 v^{14}+11114875 v^{12}+4109934 v^{10}+1293163 v^8+336754 v^6+69730 v^4+10488 v^2+977\right)
\end{array}
\end{scriptsize}
$
\vskip2mm
$
\begin{scriptsize}
\setlength{\jot}{1pt}\setlength{\arraycolsep}{1pt}
\begin{array}{ll}
R_4(v)=&-4 v^{27} \left(1-v^2\right)^{14} \left(v^2+1\right)^5 \left(49 v^{72}+2108 v^{70}+43080 v^{68}+549410 v^{66}+4790749 v^{64}+30058913 v^{62}+140688374 v^{60}+506164273 v^{58}\right.\\
&+1441227651 v^{56}+3348447760 v^{54}+6554794879 v^{52}+11146426321 v^{50}+16881715807 v^{48}+23145111240 v^{46}+28925854312 v^{44}\\
&+32964513492 v^{42}+34167708183 v^{40}+32133938929 v^{38}+27390632399 v^{36}+21164602226 v^{34}+14837233845 v^{32}+9446315545 v^{30}\\
&+5465146902 v^{28}+2873058985 v^{26}+1370892181 v^{24}+592333450 v^{22}+230873909 v^{20}+80720917 v^{18}+25125213 v^{16}+6891342 v^{14}\\
&\left.+1643932 v^{12}+335010 v^{10}+56912 v^8+7785 v^6+805 v^4+54 v^2+2\right)
\end{array}
\end{scriptsize}
$
\vskip2mm
$
\begin{scriptsize}
\setlength{\jot}{1pt}\setlength{\arraycolsep}{1pt}
\begin{array}{ll}
R_5(v)=&4 v^{36} \left(1-v^2\right)^{18} \left(v^2+1\right)^6 \left(115 v^{52}+4532 v^{50}+81255 v^{48}+870614 v^{46}+6238707 v^{44}+31899460 v^{42}+121699697 v^{40}+358454914 v^{38}\right.\\
&+837927626 v^{36}+1591222262 v^{34}+2504793634 v^{32}+3325564152 v^{30}+3777879338 v^{28}+3713876892 v^{26}+3186553262 v^{24}\\
&+2401029808 v^{22}+1595321547 v^{20}+936772628 v^{18}+486321047 v^{16}+222799298 v^{14}+89693715 v^{12}+31463008 v^{10}+9491169 v^8\\
&\left.+2402110 v^6+491208 v^4+73986 v^2+6880\right)
\end{array}
\end{scriptsize}
$
\vskip2mm
$
\begin{scriptsize}
\setlength{\jot}{1pt}\setlength{\arraycolsep}{1pt}
\begin{array}{ll}
R_6(v)=&-8 v^{27} \left(1-v^2\right)^{13} \left(v^2+1\right)^5 \left(45 v^{78}+2095 v^{76}+47359 v^{74}+677268 v^{72}+6817137 v^{70}+51164363 v^{68}+297063243 v^{66}+1371216276 v^{64}\right.\\
&+5149272306 v^{62}+16063002773 v^{60}+42426906359 v^{58}+96514459283 v^{56}+191854018167 v^{54}+337160685042 v^{52}+528528289531 v^{50}\\
&+744004161789 v^{48}+945310852132 v^{46}+1088433942563 v^{44}+1139413253221 v^{42}+1087484165439 v^{40}+948641144403 v^{38}\\
&+758021273806 v^{36}+555951360335 v^{34}+374938429891 v^{32}+232891776318 v^{30}+133416822399 v^{28}+70561418381 v^{26}\\
&+34470900225 v^{24}+15552665929 v^{22}+6474466362 v^{20}+2481641253 v^{18}+872865395 v^{16}+280231855 v^{14}+81507818 v^{12}\\
&\left.+21230952 v^{10}+4872841 v^8+958540 v^6+154859 v^4+18678 v^2+1353\right)
\end{array}
\end{scriptsize}
$
\vskip2mm
$
\begin{scriptsize}
\setlength{\jot}{1pt}\setlength{\arraycolsep}{1pt}
\begin{array}{ll}
R_7(v)=&2 v^{18} \left(1-v^2\right)^9 \left(v^2+1\right)^2 \left(38 v^{102}+2690 v^{100}+77906 v^{98}+1321893 v^{96}+15204047 v^{94}+128122734 v^{92}+830833826 v^{90}\right.\\
&+4292310349 v^{88}+18154043687 v^{86}+64331892131 v^{84}+195024248846 v^{82}+515746798173 v^{80}+1211990456294 v^{78}\\
&+2572979757047 v^{76}+4997533902047 v^{74}+8950398979633 v^{72}+14831079214352 v^{70}+22752470760842 v^{68}+32301117840063 v^{66}\\
&+42414865480784 v^{64}+51511363589878 v^{62}+57895304693808 v^{60}+60305624913172 v^{58}+58342984470106 v^{56}+52567229302146 v^{54}\\
&+44240809346010 v^{52}+34881758319792 v^{50}+25837823561098 v^{48}+18026017972892 v^{46}+11870833196234 v^{44}+7391830806862 v^{42}\\
&+4357199171850 v^{40}+2432450602770 v^{38}+1285696085386 v^{36}+642782259412 v^{34}+303479729109 v^{32}+135033780867 v^{30}\\
&+56486555930 v^{28}+22154649538 v^{26}+8122999705 v^{24}+2775291119 v^{22}+880478267 v^{20}+258402074 v^{18}+69852489 v^{16}\\
&\left.+17300406 v^{14}+3893639 v^{12}+786395 v^{10}+139397 v^8+20896 v^6+2482 v^4+195 v^2+6\right)   
\end{array}
\end{scriptsize}
$
\vskip2mm
$
\begin{scriptsize}
\setlength{\jot}{1pt}\setlength{\arraycolsep}{1pt}
\begin{array}{ll}
R_8(v)=&-4 v^{18} \left(1-v^2\right)^8 \left(v^2+1\right)^2 \left(9 v^{108}+402 v^{106}+9291 v^{104}+148334 v^{102}+1834035 v^{100}+18436088 v^{98}+153528165 v^{96}+1067107912 v^{94}\right.\\
&+6228058445 v^{92}+30771048294 v^{90}+130023959613 v^{88}+475151363552 v^{86}+1518062344822 v^{84}+4283257835180 v^{82}\\
&+10771287221805 v^{80}+24341178497700 v^{78}+49786112500029 v^{76}+92721601675056 v^{74}+158016239343581 v^{72}+247425956653936 v^{70}\\
&+357222601275580 v^{68}+477020973170522 v^{66}+590813281351395 v^{64}+680365680918064 v^{62}+730032137781654 v^{60}\\
&+731236620085836 v^{58}+684875663825102 v^{56}+600713414209744 v^{54}+494151118767024 v^{52}+381778789628840 v^{50}\\
&+277423821530342 v^{48}+189877649784600 v^{46}+122578889251559 v^{44}+74742760900738 v^{42}+43103767998145 v^{40}\\
&+23540034796206 v^{38}+12188872588015 v^{36}+5990316039836 v^{34}+2796764082671 v^{32}+1241234644808 v^{30}+523792456965 v^{28}\\
&+210120284142 v^{26}+80059059565 v^{24}+28924969344 v^{22}+9884336738 v^{20}+3182862396 v^{18}+960885861 v^{16}+270048388 v^{14}\\
&\left.+70006251 v^{12}+16515356 v^{10}+3485599 v^8+638020 v^6+97690 v^4+11138 v^2+849\right)
\end{array}
\end{scriptsize}
$
\vskip2mm
$
\begin{scriptsize}
\setlength{\jot}{1pt}\setlength{\arraycolsep}{1pt}
\begin{array}{ll}
R_9(v)=&-4 v^9 \left(1-v^2\right)^4 \left(-4 v^{128}-244 v^{126}-7365 v^{124}-139111 v^{122}-1818740 v^{120}-17469883 v^{118}-127663612 v^{116}-719958318 v^{114}\right.\\
&-3096661450 v^{112}-9512015692 v^{110}-14936878245 v^{108}+39889673215 v^{106}+435433989765 v^{104}+2112071052349 v^{102}\\
&+7557040065336 v^{100}+22176095282056 v^{98}+55809060918671 v^{96}+123567620286698 v^{94}+244922797491251 v^{92}\\
&+440265724003647 v^{90}+725056493721384 v^{88}+1102935371482521 v^{86}+1560128081733519 v^{84}+2063345885954465 v^{82}\\
&+2562281286856739 v^{80}+2996573075393827 v^{78}+3306477589653816 v^{76}+3445545020950691 v^{74}+3392166924285673 v^{72}\\
&+3155817266842722 v^{70}+2775076958874760 v^{68}+2307639616397643 v^{66}+1815970805693180 v^{64}+1353719083630132 v^{62}\\
&+957088342389309 v^{60}+642648363881471 v^{58}+410428487072540 v^{56}+249701633747747 v^{54}+144951151715494 v^{52}\\
&+80415624402956 v^{50}+42704237471148 v^{48}+21740630081494 v^{46}+10625253851487 v^{44}+4990831123645 v^{42}+2254998813535 v^{40}\\
&+980547815429 v^{38}+410359860636 v^{36}+165230528414 v^{34}+63969291831 v^{32}+23793526862 v^{30}+8495103277 v^{28}+2908669977 v^{26}\\
&+954069264 v^{24}+299397239 v^{22}+89717183 v^{20}+25601249 v^{18}+6920251 v^{16}+1761243 v^{14}+416822 v^{12}+90801 v^{10}\\
&\left.+17747 v^8+3044 v^6+428 v^4+47 v^2+2\right)
\end{array}
\end{scriptsize}
$
\vskip2mm
so that
\vskip2mm
$
\begin{scriptsize}
\setlength{\jot}{0pt}\setlength{\arraycolsep}{1pt}
\begin{array}{ll}
P_{1,SM(3,2)}(v)=&2 v^{74}+35 v^{72}+8 v^{71}+304 v^{70}+316 v^{69}+1343 v^{68}+6260 v^{67}+2234 v^{66}+66184 v^{65}-14398 v^{64}+406596 v^{63}-88904 v^{62}\\
&+1706488 v^{61}-193134 v^{60}+5581140 v^{59}-130646 v^{58}+14706492 v^{57}+246715 v^{56}+31954360 v^{55}+581552 v^{54}\\
&+58165388 v^{53}+107906 v^{52}+90927084 v^{51}-271882 v^{50}+123757956 v^{49}-102481 v^{48}+146498180 v^{47}+406866 v^{46}\\
&+151653144 v^{45}+186192 v^{44}+137665620 v^{43}-407354 v^{42}+111498240 v^{41}-891319 v^{40}+80748120 v^{39}-347946 v^{38}\\
&+52940500 v^{37}+96694 v^{36}+31684044 v^{35}+399078 v^{34}+17506960 v^{33}+383067 v^{32}+8905612 v^{31}+234956 v^{30}\\
&+4206344 v^{29}+29634 v^{28}+1866492 v^{27}-40674 v^{26}+792444 v^{25}-66831 v^{24}+319336 v^{23}-54404 v^{22}+118756 v^{21}\\
&-36518 v^{20}+42132 v^{19}-17910 v^{18}+12964 v^{17}-7755 v^{16}+3628 v^{15}-2806 v^{14}+872 v^{13}-1124 v^{12}+156 v^{11}-304 v^{10}\\
&+8 v^9-152 v^8-18 v^6-17 v^4-1
\end{array}
\end{scriptsize}
$
\vglue2mm
{\parindent0pt whose real root between $0$ and $1$ is $v_c\approx 0,5924$ corresponding to the critical temperature $T_c\approx 1,468$. $v_c$ is also the single real root between $0$ and $1$ of}
\vskip2mm
$
\begin{scriptsize}
\setlength{\jot}{1pt}\setlength{\arraycolsep}{1pt}
\begin{array}{ll}
Q^-_{1,SM(3,2)}(v)=&-\sqrt{2} v^{37}+7 v^{36}-21 \sqrt{2} v^{35}+\left(73-2 \sqrt{2}\right) v^{34}+\left(6-111 \sqrt{2}\right) v^{33}+\left(296-58 \sqrt{2}\right) v^{32}+\left(40-371 \sqrt{2}\right) v^{31}\\
&+\left(1048-484 \sqrt{2}\right) v^{30}-2 \left(13+486 \sqrt{2}\right) v^{29}+\left(2693-1459 \sqrt{2}\right) v^{28}+\left(192-1859 \sqrt{2}\right) v^{27}+\left(5154-2611\sqrt{2}\right) v^{26}\\
&+\left(274-2792 \sqrt{2}\right) v^{25}+\left(6445-3342 \sqrt{2}\right) v^{24}-2 \left(171+1513 \sqrt{2}\right) v^{23}+\left(6122-3057 \sqrt{2}\right) v^{22}-2\left(89+1473 \sqrt{2}\right) v^{21}\\
&+\left(4515-2351 \sqrt{2}\right) v^{20}-4 \left(60+491 \sqrt{2}\right) v^{19}+\left(3242-1524 \sqrt{2}\right) v^{18}+\left(18-1264 \sqrt{2}\right) v^{17}+\left(1791-730 \sqrt{2}\right) v^{16}\\
&+\left(28-571 \sqrt{2}\right) v^{15}+\left(844-442 \sqrt{2}\right) v^{14}+\left(138-296 \sqrt{2}\right) v^{13}+\left(311-237 \sqrt{2}\right)v^{12}+\left(64-107 \sqrt{2}\right) v^{11}\\
&+\left(138-71 \sqrt{2}\right) v^{10}+\left(22-56 \sqrt{2}\right) v^9+\left(59-14 \sqrt{2}\right) v^8+\left(2-16 \sqrt{2}\right)v^7\\
&+\left(18-\sqrt{2}\right) v^6+\left(2-9 \sqrt{2}\right) v^5+\left(10-\sqrt{2}\right) v^4-\sqrt{2} v^3+v^2-\sqrt{2} v+1
\end{array}
\end{scriptsize}
$
\vskip2mm
The polynom $Q^+_{1,SM(3,2)}(v)$ is obtained by the substitution $-\sqrt{2}\rightarrow +\sqrt{2}$ in $Q^-_{1,SM(3,2)}(v)$

\end{document}